%% file: main.tex
\documentclass[11pt, a4paper, logo, onecolumn,copyright,most]{tmlrmel}

\usepackage[authoryear, sort&compress, round]{natbib}
\makeatletter
\providecommand{\captionlabel}{\@ifnextchar\bgroup\captionlabel@i\captionlabel@ii}
\newcommand{\captionlabel@i}[1]{}
\newcommand{\captionlabel@ii}{}
\makeatother

\title{Combating Data Laundering in LLM Training}

\usepackage{algorithm}
\usepackage{algpseudocode}
\usepackage[utf8]{inputenc} % allow utf-8 input
\usepackage[T1]{fontenc}    % use 8-bit T1 fonts
\usepackage{hyperref}       % hyperlinks
\usepackage{url}            % simple URL typesetting
\usepackage{booktabs}       % professional-quality tables
\usepackage{nicefrac}       % compact symbols for 1/2, etc.
\usepackage{microtype}      % microtypography
\usepackage{amsmath}
\usepackage{graphicx}
\usepackage{multicol}
\usepackage{siunitx}
\usepackage{array}
\usepackage{makecell}
\usepackage[nameinlink]{cleveref}
\usepackage{bbm}
\usepackage{multirow}
\usepackage{subfig}
\usepackage{soul}
\usepackage{subcaption}
\usepackage{wrapfig}
\usepackage{blindtext}
\usepackage{tablefootnote}
\usepackage{amsfonts}
\usepackage[flushleft]{threeparttable}
\usepackage{colortbl}
\usepackage{adjustbox}
\usepackage{xcolor}
\usepackage{bbding}
\usepackage{lipsum}
\usepackage[toc,page,header]{appendix}
\theoremstyle{plain}

\theoremstyle{definition}

\theoremstyle{remark}

\usepackage{thm-restate}
\usepackage{adjustbox}

\definecolor{SteelBlue}{RGB}{70, 130, 180}
\definecolor{lg}{gray}{0.9}
\definecolor{dg}{RGB}{0,150,0}
\definecolor{dr}{RGB}{139,0,0}
\definecolor{stage1blue}{HTML}{5da0b6}
\definecolor{stage2orange}{HTML}{76dab3}
\definecolor{SDRgreen}{HTML}{ddd872}

\newtcolorbox{takeaway}{
    enhanced,
    breakable,                
colback=SteelBlue!8, 
    frame hidden,             
    borderline west={2pt}{0pt}{black!70}, 
    sharp corners,            
    fontupper=\small\itshape, 
    left=10pt, right=10pt,
    before skip=10pt, after skip=10pt
}

\input{commands}
\input{math_commands}

\author[$\diamondsuit$]{Muxing Li$^{*}$}
\author[$\diamondsuit$]{Zesheng Ye$^{*}$}
\author[$\heartsuit$]{Sharon Li}
\author[$\diamondsuit$]{Feng Liu}

\affil[ \hspace{-0.2em}]{$\diamondsuit$ University of Melbourne
}
\affil[ \hspace{-0.2em}]{$\heartsuit$ University of Wisconsin-Madison}

\correspondingauthors{fengliu.ml@gmail.com}
\equalcontributions
\acceptedbyworkshop
\newcommand{\ourmethod}{\texttt{\textbf{SDR}}}

\begin{document}

\begin{abstract}
Post-hoc unauthorized-training data detection for {\em large language models} (LLMs) typically assumes a query-with-originals regime: rights holders query a {\em target} LLM with raw proprietary data and assess whether the model assigns them stronger memorization-based detection signals, e.g., higher confidence or lower loss, than held-out non-training reference texts. We show that this regime becomes brittle under data laundering, where the {\em target} LLM is trained on semantics-preserving but stylistically or structurally transformed surrogates of proprietary data to obfuscate provenance. Since training-time exposure occurs in the laundered form, memorization signals may no longer appear on the originals, collapsing the candidate-reference signal separation that standard detectors rely on. We counter this threat by studying laundering-aware detection with raw proprietary data, a held-out reference corpus, and query access to the {\em target} LLM, while the laundering transformation is undisclosed. Since exact recovery of the laundered corpus is infeasible, we infer a detection-useful synthesis process via an auxiliary LLM that maps originals into training-like queries. To make this search tractable, we introduce Synthesis Data Reversion (\ourmethod), which constrains the unbounded space of natural-language transformations through a goal-details abstraction: a high-level transformation goal, e.g., "lyrical rewriting", and fine-grained details, e.g., "with vivid imagery". \ourmethod~identifies the most likely goal and iteratively refines details so synthesized queries elicit stronger target-model detection signals. Evaluated on the MIMIR benchmark against diverse laundering practices and target LLM families (Pythia, Llama2, and Falcon), \ourmethod~consistently restores detection signals, offering a practical auditing layer against data laundering.
\end{abstract}

\maketitle

\input{sections/introduction}
\newpage
\bibliography{main}
\input{sections/appendix}

% \clearpage
% \input{sections/appendix}
\clearpage

\end{document}

%% file: commands.tex
\usepackage{xspace}
\newcommand{\draftonly}[1]{#1}
\newcommand{\eat}[1]{}
\renewcommand{\draftonly}[1]{}
\definecolor{darkgreen}{RGB}{0, 102, 0}

\crefformat{section}{\S#2#1#3}

%% file: math_commands.tex
\usepackage{amsmath,amsfonts,bm}

\def\eqref#1{equation~\ref{#1}}
\def\1{\bm{1}}

\DeclareMathAlphabet{\mathsfit}{\encodingdefault}{\sfdefault}{m}{sl}
\SetMathAlphabet{\mathsfit}{bold}{\encodingdefault}{\sfdefault}{bx}{n}

%% file: sections/introduction.tex
\section{Introduction}

\emph{Large language models} (LLMs) can generate fluent and stylistically diverse text, driving adoption in medicine~\citep{liu2025generalist}, education~\citep{yan2024practical}, and other high-stakes domains. These capabilities require training on large, high-quality corpora~\citep{wang2025comprehensive}, whose collection and use are often constrained by privacy and copyright~\citep{li2023privacy}. A central compliance question is therefore whether a deployed LLM was trained on copyrighted or sensitive data without authorization.

\begin{figure}[!t]
    \centering
    \includegraphics[width=0.9\linewidth]{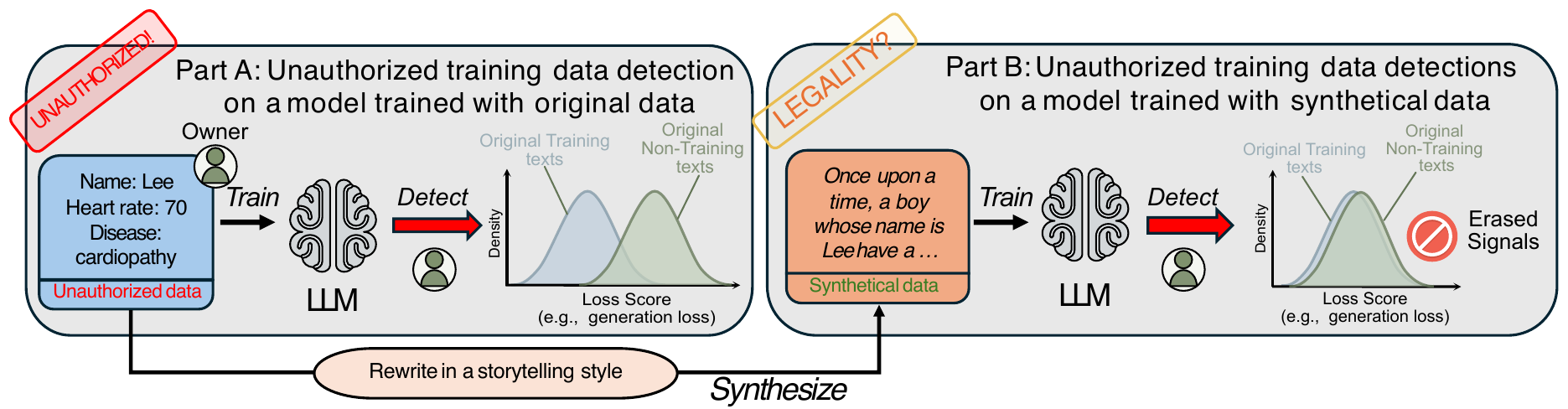}
    % \vspace{-2em}
    \caption{\small 
Illustration of how data laundering undermines existing unauthorized training data detectors. When unauthorized data is directly used for training, LLMs tend to memorize the unauthorized training data. \textcolor{blue!32}{Original training samples} exhibit lower loss than \textcolor{green!50!black}{original non-training samples}, as shown in Part A. The log-likelihood distributions of training and non-training samples diverge clearly, enabling identification. However, when trained on laundered synthetic data, as shown in Part B, the distributions of original training samples and original non-training samples no longer diverge, preventing reliable identification.
}
    \label{fig:mot}
    \vspace{-1.5em}
\end{figure}

Post-hoc \emph{unauthorized training data detection} addresses this question by querying a \emph{target} LLM with proprietary candidate texts and comparing their scores, such as loss~\citep{zhang2024min} or calibrated confidence~\citep{xie2024recall}, against those of a held-out non-training reference corpus, following~\citet{carlini2022membership}. Since LLMs can memorize training data~\citep{li2025membership}, training samples often receive lower loss or higher confidence than non-training samples. This score gap enables reliable detection when the target model was trained on the original proprietary texts (Figure~\ref{fig:mot}~(Part A)); mainstream methods perform well in this ``query with originals'' regime (Table~\ref{tab:1_compare_orig_syn_detection_performance}, ``Orig.'' columns).

This regime implicitly assumes that, if the target LLM used the rights owner's data, it was exposed to the data in its original textual form. Natural language, however, can be extensively rewritten while preserving its underlying semantics~\citep{barzilay2001extracting, bhagat2013paraphrase}. Such rewrites can be produced through human writing, programmatic paraphrase, back-translation~\citep{dolan2005automatically, bannard2005paraphrasing}, or LLM-based synthesis~\citep{witteveen2019paraphrasing, liu2024monotonic}. If a model provider trains only on these transformed surrogates, the target LLM may memorize the surrogates rather than the originals. As a result, querying with the originals no longer yields a reliable score gap (Figure~\ref{fig:mot}~(Part B)). Empirically, when a Llama-2 model~\citep{touvron2023llama} is trained on Wikipedia articles rewritten into a lyrical style, standard detection methods applied to the originals perform close to random chance (Table~\ref{tab:1_compare_orig_syn_detection_performance}, ``Syn.'' columns).

We call this evasion mechanism \emph{data laundering}: the model provider transforms proprietary data into semantics-preserving surrogates, trains on the surrogates, and can later claim that the originals were never used. This creates practical risks for copyright and privacy auditing. Data rights owners usually have only black-box API access to the target LLM, and providers rarely disclose preprocessing pipelines, data sources, or training artifacts. The laundering transformation, if it exists, is therefore hidden from the auditor. We ask:

\begin{takeaway}
{How can a data rights owner detect unauthorized data use from black-box model access when the data has been laundered through an unknown transformation?}
\end{takeaway}

Our key idea is to synthesize queries that resemble the laundered data seen during training. If the target model was trained on transformed surrogates, then querying it with similar surrogates should restore the memorization-based score gap. Since the true transformation is unknown, we use an \emph{auxiliary} LLM as a generator~\citep{liang2024controllable} and search for a prompt that maps original texts into training-like rewrites. The challenge is that the space of possible rewriting prompts is too large to search directly.

To make this search tractable, we model the unknown transformation into a coarse \emph{goal} and fine-grained \emph{details}. The goal captures the dominant register or structural shift, such as rewriting into lyrics, storytelling, or instructions. The details specify additional constraints, such as imagery, voice, formatting, or rhyme density. This goal-detail decomposition follows the structure of common prompt templates~\citep{mao2025prompts}: the goal defines the main rewriting direction, while the details refine the generated surrogates. We use this decomposition only as a search strategy for producing training-like queries, not as a claim that every real laundering pipeline must follow this exact form.\footnote{A register is a situational variety of language shaped by purpose, audience, and medium; examples include news, academic prose, instructions, and lyrics~\citep{agha2004registers}.}

Building on this decomposition, we introduce \emph{synthesis data reversion}~(\ourmethod), a two-stage method for laundering-aware detection. First, \ourmethod~identifies a coarse laundering goal by searching over a bounded set of linguistic registers and selecting the one whose synthesized queries best match the target model's black-box behavior. Second, \ourmethod~refines the selected goal by using target-model feedback to infer missing details and update the synthesis prompt. The final prompt produces training-like surrogates for both the proprietary candidate set and the held-out reference set, enabling off-the-shelf detectors, such as Recall~\citep{xie2024recall}, to be applied without modification.

We evaluate \ourmethod~on MIMIR benchmark~\citep{deng2023investigating} across target LLM families including Pythia~\citep{biderman2023pythia}, Llama-2~\citep{touvron2023llama}, and Falcon~\citep{almazrouei2023falcon}, as well as auxiliary LLM choices including DeepSeek~\citep{liu2024deepseek}, GPT-4o~\citep{roumeliotis2023chatgpt}, and Claude~\citep{wu2023comparative}. Across diverse simulated laundering procedures, \ourmethod~consistently strengthens standard detection methods, and ablation studies show that both stages contribute to the improvement.

\noindent
\textbf{Contributions.}
This paper makes three contributions. First, we formulate data laundering-aware post-hoc unauthorized training data detection for black-box LLMs. Second, we introduce a goal-detail decomposition that turns the open-ended transformation search into a tractable synthesis problem. Third, we propose \ourmethod, a two-stage method that restores the effectiveness of standard detection methods under data laundering. Together, these results provide a practical auditing approach for data rights holders and highlight data laundering as an emerging risk for LLM governance.

\begin{table}[t]
\centering
\caption{\small Performance of unauthorized training data detection on Llama-2~\citep{touvron2023llama} models fine-tuned with either the original MIMIR-Wikipedia~\citep{deng2023investigating} dataset (Orig.) or its laundered version (Syn.) generated by GPT-4o~\citep{hurst2024gpt} using the prompt “rewrite in a lyrical style, ensuring the imagery is vivid". Evaluation metrics are defined in Section~\ref{sec:experiment}.}
\vspace{1mm}
\footnotesize
\begin{tabular}{lcccccc} % ← 恢复为 7 列：l + 6 个 c
\toprule
\multirow{2}{*}{Methods}
& \multicolumn{2}{c}{AUC}
& \multicolumn{2}{c}{ASR}
& \multicolumn{2}{c}{TPR@5\%} \\
\cmidrule(lr){2-3}\cmidrule(lr){4-5}\cmidrule(lr){6-7}
& Orig. & Syn.
& Orig. & Syn.
& Orig. & Syn. \\
\midrule
Loss~\citep{yeom2018privacy} & 1.000 & 0.539 & 1.000 & 0.565 & 1.000 & 0.040 \\
\midrule
Ref~\citep{carlini2022membership}        & 0.971 & 0.603 & 0.920 & 0.610 & 0.850 & 0.100 \\
\midrule
Zlib~\citep{carlini2021extracting} & 1.000 & 0.521 & 1.000 & 0.535 & 1.000 & 0.080 \\
\midrule
Min-K~\citep{shi2023detecting}     & 1.000 & 0.563 & 1.000 & 0.575 & 1.000 & 0.040 \\
\midrule
% Min-K++~\citep{zhang2024min}    & 1.000 & 0.547 & 1.000 & 0.575 & 1.000 & 0.040 \\
Recall~\citep{xie2024recall}     & 0.999 & 0.558 & 0.995 & 0.565 & 1.000 & 0.000 \\
\bottomrule
\end{tabular}
% \end{small}
\label{tab:1_compare_orig_syn_detection_performance}
\vspace{-2em}
\end{table}

\vspace{-0.5em}\section{Unauthorized Data Detection: Progress and Challenge}
\label{sec:related work}

\vspace{-0.5em}Verifying the provenance of data used to train LLMs is a central problem for trustworthy AI~\citep{li2023trustworthy}. Existing solutions can be broadly divided into proactive defenses and post-hoc detection.

\noindent
\textbf{Progress in unauthorized data detection.}
Proactive defenses aim to prevent or trace misuse before or during training. Data watermarking embeds signals, such as stylistic patterns, into training data so that later model outputs can reveal provenance~\citep{liang2024watermarking}. Differential privacy~\citep{dwork2008differential} provides formal guarantees against memorization by adding calibrated noise during training, but it often reduces model utility and is rarely adopted in LLM training. More importantly, these approaches require action before training and cannot audit already deployed models whose data sources are undisclosed.

Post-hoc detection instead asks whether a deployed model has used specific data after the model has been trained. This setting is especially relevant for LLM auditing, where data rights owners often have only black-box access to the target model. Most post-hoc methods build on membership-inference ideas~\citep{shokri2017membership}: training samples tend to induce different model behaviors from unseen samples because overparameterized models can memorize their training data~\citep{li2025membership}. In LLMs, this difference is commonly measured through intrinsic signals such as loss, likelihood, or calibrated confidence~\citep{carlini2021extracting, carlini2022membership, ye2024data, shi2023detecting, zhang2024min, xie2024recall}. These methods provide practical tools for detecting unauthorized data use and remain effective in the standard setting where the queried texts match the texts seen during training.

\noindent
\textbf{Data laundering challenges post-hoc detection.}
Existing post-hoc detectors are designed and evaluated under the ``query with originals'' regime: the rights holder queries the target LLM with original proprietary texts and compares their scores against those of a non-member reference corpus. This regime breaks down when the target model was trained not on the originals, but on semantic-preserving surrogates. Such surrogates can be produced through paraphrasing, back-translation~\citep{barzilay2001extracting, bannard2005paraphrasing}, register or style transfer, or large-scale LLM-based rewriting~\citep{witteveen2019paraphrasing, zeleke2025human}.

We refer to this evasion mechanism as \emph{data laundering}. Under data laundering, the target model may memorize the transformed surrogates rather than the original texts. As a result, original member samples can lose their score advantage over non-members, causing the candidate-reference separation used by standard detectors to collapse. The challenge is further amplified by the opacity of real-world LLM deployment: data rights owners usually do not know whether laundering occurred, what transformation was used, or what surrogate texts appeared in training.

% \noindent
% \textbf{From exact recovery to detection-useful synthesis.}
% A direct solution would be to recover the exact laundered training texts and query the target model with them. This is unrealistic under black-box access, because the laundering transformation is hidden and the space of possible semantics-preserving rewrites is too large for brute-force search. Therefore, the goal should not be exact recovery of the hidden transformation. Instead, it is sufficient to find a \emph{detection-useful} synthesis process: a prompt-controlled transformation that maps both proprietary candidate texts and held-out reference texts into queries close enough to the target model's laundered training distribution.

% If such synthesized queries restore the score separation between the candidate and reference sets, existing off-the-shelf detectors can be reused without changing their scoring rules. This motivates the formulation in the next section: we search for a synthesis prompt that maximizes the detection performance induced by synthesized candidate and reference corpora.

\vspace{-0.5em}\section{Reverse-engineering the Laundering Transformation}
\label{sec:preliminary}

\vspace{-0.5em}The challenge above suggests a direct but unrealistic solution: recover the exact laundered training texts and query the target model with them. Under black-box access, however, exact recovery is infeasible because the laundering transformation is hidden and the space of semantics-preserving rewrites is too large for brute-force search. We therefore seek a weaker but sufficient goal: a \emph{detection-useful} synthesis process that maps both proprietary candidate texts and held-out reference texts into queries close enough to the target model's laundered training distribution to restore candidate-reference score separation. We then formalize this objective below.

\textbf{Problem setup.}
Let \(M_t\) denote the target LLM, trained on an unknown corpus \(D_{\rm train}\), and let \(M_a\) denote an auxiliary LLM used by the data rights owner for synthesis. The rights owner has a proprietary candidate corpus \(D_{\rm pro}\) and a held-out non-training reference corpus \(D_{\rm held}\). The target training corpus may contain laundered variants of a subset of \(D_{\rm pro}\), but the rights owner has access only to \(D_{\rm pro}\), \(D_{\rm held}\), and black-box queries to \(M_t\). The owner does not observe \(D_{\rm train}\), the laundered corpus, or the provider's preprocessing pipeline.

A synthesis prompt \(p\), executed by \(M_a\), induces a prompt-based transformation \(\mathrm{Syn}_p(\cdot)\). Applying this transformation to the candidate and reference corpora yields two synthesized sets, \(\mathrm{Syn}_p(D_{\rm pro})\) and \(\mathrm{Syn}_p(D_{\rm held})\). These sets are then evaluated on \(M_t\) using an off-the-shelf unauthorized training data detector.

\textbf{Objective.}
Our goal is to find a prompt \(p\) that makes the synthesized candidate and reference sets maximally separable under the chosen detector:
\begin{align}   
    p^\star
    =
    \arg\max_p
    \mathrm{Perf}\big(
    M_t,
    \mathrm{Syn}_p(D_{\rm pro}),
    \mathrm{Syn}_p(D_{\rm held})
    \big),
\end{align}   
where \(\mathrm{Perf}(\cdot)\) denotes the detector's performance measure, such as AUC, ASR, or TPR at a fixed false-positive rate. A high value of \(\mathrm{Perf}\) means that the synthesized candidate texts elicit stronger training-data signals than the synthesized reference texts. In this case, \(p\) serves as a detection-useful approximation to the hidden laundering transformation: it need not recover the exact training-time surrogates, but it produces queries that restore the score separation needed by standard detectors. The detailed threat model and objective are provided in Appendix~\ref{app:threat}.

\textbf{Managing the prompt search space.}
The objective above still involves searching over natural-language prompts, whose space is effectively unbounded~\citep{zhang2025prompt}. 
We therefore restrict the search by decomposing each synthesis prompt into a coarse \emph{goal} and fine-grained \emph{details}. 
This decomposition follows common prompt-template structures~\citep{mao2025prompts} and linguistic theory~\citep{agha2004registers, myntti2025register}, where a prompt contains a core directive and additional modifiers such as context and constraints. 
In our setting, the \emph{goal} specifies the dominant transformation direction, such as rewriting the input in a lyrical style, while the \emph{details} specify supplementary requirements that refine the generated surrogate, such as vivid imagery, narrative voice, or formatting constraints.

To make the goal search finite, we ground the goal in a linguistic register taxonomy. 
Specifically, we search over 23 registers (collected in a set $R$) that cover major communicative forms~\citep{henriksson2024automatic}, using each register as a candidate coarse transformation goal. 
The remaining problem is then to select the register most compatible with the target model's black-box behavior and refine the details within that register.\footnote{We acknowledge that this taxonomy was not designed for data laundering and thus has limitations, which we discuss in Appendix~\ref{limitaion}.} 
This leads to the two-stage method in Section~\ref{sec:method}.

\begin{algorithm}[!t]
\small
\caption{Goal identification stage}
\label{alg:directive}
\begin{algorithmic}[1]
\Require Proprietary originals $D_{\mathrm{pro}}$, held-out data $D_{\mathrm{held}}$, target model $M_t$, auxiliary LLM $M_a$, set of 23 registers $R$, sample size $n$ and $m$
\Ensure the register $r^\ast \in R$ that is closely aligned with the laundering directive.

\State \textbf{---Constructing opening templates---}
\ForAll{$r \in R$}
  \State $\text{Standard-prompt}_r \gets M_a(\text{``Give me a prompt that can transfer text into register $r$"})$
\State $O_r \gets \{\, \text{The first sentence of }M_a (\text{Standard-Prompt}_r, s) \mid s \in \text{UniformSample} (D_{\mathrm{pro}}, n) \,\}$
  \State $\text{T}_r \gets M_a(\text{``Extract a common template."},O_r)$
\EndFor

\State \textbf{---Scoring via continuation confidence---}
\ForAll{$r \in R$}
  \State $S \gets \{\, \text{The first sentence of } s \mid s \in \text{UniformSample}(D_{\mathrm{pro}}, m) \,\}$
  \State $X_r \gets \{\, M_a(\text{``Rewrite } x \text{ as } \text{T}_r") \mid x \in S\,\}$
  \For{$j \gets 1$ \textbf{to} $m$}
    \State $c_j \gets \text{Average next token confidence of }M_t(X_r[j])$
  \EndFor
  \State $\mathrm{Conf}(r) \gets \frac{1}{m}\sum_j c_j$
\EndFor
\State $C \gets$ top-5 registers with largest $\mathrm{Conf}(r)$

\State \textbf{---Selecting best register---}
\ForAll{$r \in C$}
\State $\mathrm{Syn_r} \gets \{\, M_a(\text{Standard-prompt}_r,d) \mid d \in D_{\mathrm{pro}} \cup D_{\mathrm{held}} \,\}$
% \State $\mathrm{Perf}_r \gets \text{\State Perform unauthorized training data detection on $M_t$ using queries from $\mathrm{Syn_r}$}$
% \State $\mathrm{Perf}_r \gets \frac{\text{Loss}(\mathrm{Syn_r}, M_t)+ \text{Ref}(\mathrm{Syn_r}, M_t)+ \text{Zlib}(\mathrm{Syn_r}, M_t)+ \text{Min-K}(\mathrm{Syn_r}, M_t)+ \text{Recall}(\mathrm{Syn_r}, M_t)}{5}$
\State $\mathrm{Perf}_r \gets$ Unauthorized training data detection on $M_t$ using $\mathrm{Syn_r}$
\EndFor
\State $r^\ast \gets \arg\max_{r \in C} \mathrm{Perf}_r$
\State \Return $r^\ast$, $\text{Standard-prompt}_{r^\ast}$ 
\end{algorithmic}
\end{algorithm}

\section{Synthesis Data Reversion}
\label{sec:method}

We propose \emph{synthesis data reversion} (\ourmethod), a two-stage method for searching the prompt \(p\) defined in Section~\ref{sec:preliminary}. 
Given the goal-detail decomposition, \ourmethod~first identifies a coarse rewriting goal and then refines the missing details within that goal. 
The output is a prompt that synthesizes ``training-like'' variants of both \(D_{\mathrm{pro}}\) and \(D_{\mathrm{held}}\), so that standard unauthorized training data detectors can be applied to the synthesized corpora. 
% The proposed \ourmethod~consists of two algorithms corresponding to each stage, which will be introduced below.
The proposed \ourmethod~proceeds in two stages: goal identification and detail refinement, described in Algorithms~\ref{alg:directive} and~\ref{alg:condition}, respectively.
% consists of two algorithms corresponding to each stage, which will be introduced below.

\subsection{Goal Identification Stage}

The first stage selects the register \(r^\ast\) that is most compatible with the hidden laundering goal (Algorithm~\ref{alg:directive}). 
A direct solution would synthesize all samples in \(D_{\mathrm{pro}}\cup D_{\mathrm{held}}\) into every register and run the detector for each one. 
This is expensive because it requires many long-form rewrites. 
Instead, \ourmethod~uses opening sentences as a cheap proxy: if the true laundering process rewrites data into a particular register, then openings written in that register should look familiar to the target model and elicit high-confidence continuations. 
For example, if the hidden laundering prompt rewrites Wikipedia text into a lyrical style, an opening such as ``In the heart of \ldots'' should be more compatible with the target model than an opening written as a legal clause or a recipe. To implement this idea, the first block in Algorithm~\ref{alg:directive} is designed to construct opening templates.

\textbf{Constructing opening templates.}
Lines~2--6 of Algorithm~\ref{alg:directive} build an opening template for each register \(r\in R\). 
For each \(r\), \ourmethod~first asks the auxiliary LLM \(M_a\) to produce a \emph{Standard-prompt} that transfers text into that register (line~3). 
It then applies this standard prompt to \(n\) sampled proprietary texts and keeps only the first sentence of each rewrite (line~4). 
Finally, \(M_a\) abstracts these first sentences into a common opening template \(T_r\) (line~5). 
For instance, for the lyrical register, the sampled rewrites may begin with vivid, image-heavy openings; \(M_a\) may summarize them into a template such as ``In the heart of [abstract domain], a tale unfolds, where \ldots''. 
This template captures how the register typically begins without rewriting the full corpus (the opening templates for each register are provided in Appendix~\ref{app:template}).

\textbf{Scoring registers by continuation confidence.}
Lines~8--16 use the target model \(M_t\) to score how compatible each register is with the training-time distribution. 
For each register \(r\), \ourmethod~samples \(m\) proprietary texts and extracts their original first sentences (line~9). 
It then rewrites these first sentences according to the template \(T_r\), producing short register-conditioned queries \(X_r\) (line~10). 
Each query \(X_r[j]\) is fed into \(M_t\) where $j=1,\dots,m$, and \ourmethod~measures how confidently the target model continues it (lines~11--14). 
Specifically, let the continuation length for \(X_r[j]\) be \(L_j\), and let \(X_{r,<i}^{(j)}\) denote the rewritten opening sentence together with the first \(i-1\) generated tokens. 
We compute
\begin{align}   
\mathrm{Conf}(r)
=
\frac{1}{m}
\sum_{j=1}^{m}
\left(
\frac{1}{L_j}
\sum_{i=1}^{L_j}
\max_{w\in V}
P_{M_t}\!\left(w \mid X_{r,<i}^{(j)}\right)
\right),
\end{align}   
where \(V\) is the vocabulary, and \(P_{M_t}(w \mid X_{r,<i}^{(j)})\) denotes the probability assigned by the target model \(M_t\) to token \(w\) as the next token given the current prefix \(X_{r,<i}^{(j)}\). 
The intuition is that a register closer to the laundering goal should make \(M_t\) more confident when continuing the opening. 
\ourmethod~keeps the top-5 registers with the largest \(\mathrm{Conf}(r)\) as a small candidate set \(C\) (line~16).

\textbf{Selecting the best register.}
Lines~18--22 choose the final goal from the shortlisted registers in $C$. 
For each \(r\in C\), \ourmethod~uses the corresponding \emph{Standard-prompt} (obtained in line 3) to synthesize both the proprietary candidate corpus and the held-out reference corpus (line~19). 
It then runs the chosen unauthorized training data detector on \(M_t\) using these synthesized queries and records the resulting performance \(\mathrm{Perf}_r\) (line~20). 
The selected register \(r^\ast\) is the one that maximizes this detector performance (line~22), and \ourmethod~returns both \(r^\ast\) and its \emph{Standard-prompt} (line~23). 
Thus, stage~1 turns the open-ended goal search into a finite register search and provides the initial prompt for stage~2.

\begin{algorithm}[!t]
\small
\caption{Detail refinement stage}
\label{alg:condition}
\begin{algorithmic}[1]
\Require register $r^\ast$ and $\text{Standard-prompt}_{r^\ast}$ got from Algorithm~\ref{alg:directive} , proprietary data $D_{pro}$, held-out data $D_{\mathrm{held}}$, target model $M_t$, auxiliary LLM $M_a$, iteration budget $K$, sample size $l$
\Ensure Reversed prompt
% \State \textbf{Condition inference:}
\State $p = \text{Standard-prompt}_{r^\ast}$
\Function{ConditionInference}{$D_{\mathrm{pro}}, p, M_t, M_a$}     % 定义函数
\ForAll{$s \in \text{UniformSample}(D_{\mathrm{pro}}, l)$}
% \Forall{\textsc{Sample}(D_{\mathrm{pro}}}
  \State $\hat{s} \gets M_a(p, s)$
  \State $\tilde{s} \gets M_t(\text{the first sentence of }\hat{s})$
  % \State 
  \State $H.\textsc{Append(}M_a\text{(``Editing $p$ enables the transformation of $\hat{s}$ into $\tilde{s}$"}$))
\EndFor
\State \Return  $M_a(\text{``Extract a common prompt."},H)$     
\EndFunction

\Function{Evaluate}{$D_{pro}, D_{\mathrm{held}}, p, M_t, M_a$}     % 定义函数
\State $\mathrm{Syn_p} \gets \{\, M_a(p, x) \mid x \in D_{\mathrm{pro}} \cup \mathcal{D}_{\mathrm{held}} \,\}$
\State $\mathrm{Perf}_p \gets$ Unauthorized training data detection on $M_t$ using $\mathrm{Syn_p}$
\State \Return  $\mathrm{Perf}_p$     
\EndFunction

\For{$k \gets 0$ \textbf{to} $K-1$}
  \State $p' \gets \textsc{ConditionInference}(D_{\mathrm{pro}}, p, M_t, M_a)$
  \State $\mathrm{score}_{\text{new}} \gets \textsc{Evaluate}(D_{\mathrm{pro}}, D_{\mathrm{held}}, p', M_t, M_a)$, $\mathrm{score}_{\text{old}} \gets \textsc{Evaluate}(D_{\mathrm{pro}}, D_{\mathrm{held}}, p,  M_t, M_a)$
  \If{$\mathrm{score}_{\text{new}} > \mathrm{score}_{\text{old}}$}
    \State $p \gets p'$
  \EndIf
\EndFor
\State \Return $p$
\end{algorithmic}
\end{algorithm}

\subsection{Detail Refinement Stage}
The selected register gives only a coarse transformation direction. 
For example, knowing that the laundering goal is ``lyrical rewriting'' does not specify whether the rewrite should use vivid imagery, first-person narration, rhyme, short lines, or a particular narrative voice. 
Such missing details can matter because the target model was trained on a specific surrogate distribution, not merely on any text from the same broad register. 
The second stage therefore refines the prompt returned by Algorithm~\ref{alg:directive} using feedback from \(M_t\) (Algorithm~\ref{alg:condition}).

\textbf{Initializing the prompt.}
Line~1 of Algorithm~\ref{alg:condition} initializes the current prompt \(p\) as \(\text{Standard-prompt}_{r^\ast}\), the standard prompt associated with the selected register. 
This prompt already rewrites in the coarse target direction, but it may miss the fine-grained details to produce detection-useful queries.

\textbf{Inferring candidate refinements from target continuations.}
Lines~2--9 define \textsc{ConditionInference}. 
For each sampled proprietary text \(s\), \ourmethod~first synthesizes a surrogate \(\hat{s}=M_a(p,s)\) using the current prompt \(p\) (line~4). 
It then feeds the first sentence of \(\hat{s}\) into the target model \(M_t\) and obtains a continuation \(\tilde{s}\) (line~5). 
The first sentence serves as a trigger: if the current prompt is close to the training-time style, the target model's continuation may reveal stylistic or structural properties of the laundered data. 
\ourmethod~then asks \(M_a\) to describe how \(p\) should be edited so that \(\hat{s}\) better matches \(\tilde{s}\) (line~6). 
For example, if \(\hat{s}\) is a generic lyrical rewrite but \(\tilde{s}\) continues with denser imagery and stronger rhythm, \(M_a\) may propose adding details such as ``use vivid imagery'' or ``make the rhythm flow naturally.'' 
After collecting such suggestions over \(l\) samples, \ourmethod~asks \(M_a\) to distill them into a single revised prompt \(p'\) (line~8).

\textbf{Evaluating a refinement.}
Lines~10--14 define \textsc{Evaluate}. 
Given a prompt \(p\), \ourmethod~synthesizes both \(D_{\mathrm{pro}}\) and \(D_{\mathrm{held}}\) with \(M_a\) (line~11), runs the unauthorized training data detector on \(M_t\) using the synthesized corpora (line~12), and returns the resulting score \(\mathrm{Perf}_p\) (line~13). 
This evaluation step is essential: a refinement is useful only if it improves candidate-reference separation, not merely if it sounds closer to the target model's continuation.

\textbf{Iterative prompt refinement.}
Lines~15--21 repeat the refinement process for at most \(K\) iterations. 
At each iteration, \ourmethod~proposes a new prompt \(p'\) through \textsc{ConditionInference} (line~16), evaluates both the proposed prompt and the current prompt (line~17), and accepts the proposal only when \(\mathrm{score}_{\mathrm{new}}>\mathrm{score}_{\mathrm{old}}\) (lines~18--19). 
This acceptance rule keeps the search aligned with the objective in Section~\ref{sec:preliminary}: the prompt is refined only when it improves detection performance on synthesized candidate and reference sets. 
The final returned prompt \(p\) (line~22) is therefore a detection-useful synthesis prompt that combines a coarse register goal with refined details.

\begin{takeaway}
\emph{\textbf{Remark.} 
\ourmethod~does not aim to recover the exact laundering prompt used by the model provider. 
Instead, it seeks a \emph{detection-useful} prompt that restores candidate-reference score separation under standard detectors. 
The recovered prompt may therefore differ from the true one, so its success should be evaluated by detection performance rather than exact prompt matching. 
The next section empirically tests this criterion across different laundering prompts, datasets, target models, and auxiliary LLMs, including settings where the auditor's auxiliary LLM \(M_a\) differs from the model used for laundering. }
\end{takeaway}

% \textbf{Remark.} 
% \ourmethod~does not aim to recover the exact laundering prompt used by the model provider. 
% Instead, it seeks a \emph{detection-useful} prompt that restores candidate-reference score gap under standard detectors. 
% The recovered prompt may therefore differ from the true one, so its success should be evaluated by detection performance rather than exact prompt matching. 
% The next section empirically tests this criterion across different laundering prompts, datasets, target models, and auxiliary LLMs, including settings where the auditor's auxiliary LLM \(M_a\) differs from the model used for laundering.

% \vspace{-1em}

\section{Experiments and Results}
\label{sec:experiment}

% \subsection{Experimental Setup}

We first describe the datasets, target models, baselines, metrics, and laundering protocol.

\begin{table}[!t]
\centering
% \vspace{-1em}
\caption{
\small The average performance of each unauthorized training data
detection method across data synthesized from \textbf{different inside and outside laundering processes}. The experiment is conducted on Pythia-6.9B~\citep{biderman2023pythia}, fine-tuned on Wikipedia synthesis. The results for each prompt are provided in Appendix~\ref{App:specific}. We report the additional TPR@1\% results in Appendix~\ref{app:tpr}.
}
% \vspace{-1em}
\label{tab:main}
\renewcommand{\arraystretch}{0.9} % 调整行间距
\footnotesize
\begin{tabular}{lccc|ccc}
\toprule
\multirow{2}{*}{Method}   & \multicolumn{3}{c}{Inside registers} & \multicolumn{3}{c}{Outside registers} \\
\cmidrule(lr){2-4} \cmidrule(lr){5-7}
& AUC & ASR & TPR@5\% & AUC & ASR & TPR@5\% \\
\midrule
Recall & 64.7\% & 63.4\% & 8.9\% & 61.7\% & 61.4\% & 5.6\% \\
Recall+\ourmethod~  & \textbf{76.2\%} & \textbf{72.0\%} & \textbf{25.3\%} & \textbf{73.4\%} & \textbf{73.3\%} & \textbf{23.2\%} \\
\midrule
Loss & 63.7\% & 62.8\% & 10.7\% & 62.7\% & 62.6\% & 9.2\% \\
Loss+\ourmethod~  & \textbf{76.6\%} & \textbf{72.6\%} & \textbf{26.2\%} & \textbf{75.5\%} & \textbf{75.5\%} & \textbf{22.9\%} \\
\midrule
Ref & 68.6\% & 67.0\% & 15.0\% & 67.6\% & 65.6\% & 13.2\% \\
Ref+\ourmethod~  & \textbf{74.8\%} & \textbf{70.8\%} & \textbf{29.9\%} & \textbf{72.0\%} & \textbf{72.1\%} & \textbf{24.2\%} \\
\midrule
Zlib & 63.9\% & 63.5\% & 15.2\% & 63.6\% & 63.5\% & 13.9\% \\
Zlib+\ourmethod~  & \textbf{68.9\%} & \textbf{66.7\%} & \textbf{18.8\%} & \textbf{68.4\%} & \textbf{68.4\%} & \textbf{16.2\%} \\
\midrule
Min-K & 63.5\% & 62.6\% & 11.8\% & 64.2\% & 62.5\% & 10.5\% \\
Min-K+\ourmethod~  & \textbf{75.1\%} & \textbf{71.6\%} & \textbf{25.1\%} & \textbf{73.6\%} & \textbf{73.5\%} & \textbf{22.7\%} \\
\bottomrule
\end{tabular}
\vspace{-2em}
\end{table}

\textbf{Datasets and target models.}
We evaluate \ourmethod~on the MIMIR benchmark~\citep{deng2023investigating} (see Appendix~\ref{app:mimir} for more details), which is widely used for unauthorized training data detection. We use three subsets from MIMIR: Wikipedia, C4, and HackerNews, corresponding to encyclopedia articles, web text, and news-style discussions, respectively. Unless otherwise specified, the main experiments use MIMIR-Wikipedia and a Pythia-6.9B~\citep{biderman2023pythia} target model. To test robustness, we further evaluate different target architectures, including Pythia~\citep{biderman2023pythia}, Falcon~\citep{almazrouei2023falcon}, and Llama-2~\citep{touvron2023llama}.

\textbf{Baselines and metrics.}
We evaluate \ourmethod~as a plug-in procedure for five off-the-shelf unauthorized training data detectors: Loss~\citep{yeom2018privacy}, Ref~\citep{carlini2022membership}, Zlib~\citep{carlini2021extracting}, Min-K~\citep{shi2023detecting}, and Recall~\citep{xie2024recall}. Following prior work~\citep{carlini2022membership}, we report Area Under the Curve (AUC), Attack Success Rate (ASR), and True Positive Rate at 5\% False Positive Rate (TPR@5\%). AUC measures overall separability, ASR measures single-threshold detection accuracy, and TPR@5\% captures performance in a low-false-positive auditing regime. Detailed definitions are provided in Appendix~\ref{sec:metrics}.

\textbf{Laundering prompts and evaluation protocol.}
We simulate data laundering by rewriting original MIMIR samples into synthetic surrogates before target-model fine-tuning. We consider two prompt families. \emph{Inside-register prompts} rewrite data into one of the 23 registers used by \ourmethod, while \emph{outside-register prompts} are generated to fall outside this predefined taxonomy. The full prompt lists are provided in Appendix~\ref{app:prompts}. For each laundering prompt, the model provider synthesizes the original data, fine-tunes the target model on a subset of the synthesized data, and keeps the remaining synthesized samples as held-out references. The auditor only has access to the original candidate data, the held-out reference data, an auxiliary LLM, and black-box queries to the target model. \ourmethod~then infers a prompt, rewrites both candidate and reference sets, and runs the off-the-shelf detector on the synthesized queries. Further implementation details are provided in Appendix~\ref{sec:detail implement}.

\subsection{Main Result: \ourmethod~Restores Detection under Diverse Laundering Prompts}

Table~\ref{tab:main} reports the main results on MIMIR-Wikipedia with a Pythia-6.9B target model under both inside-register and outside-register laundering prompts. Across both prompt families, \ourmethod~consistently improves standard detectors, especially when they are evaluated by the TPR@5\%.
% Under inside-register prompts, Recall improves from 64.7\% to 76.2\% AUC and from 8.9\% to 25.3\% TPR@5\%; Loss improves from 63.7\% to 76.6\% AUC and from 10.7\% to 26.2\% TPR@5\%. Under outside-register prompts, \ourmethod~remains effective even though the true laundering prompts do not directly correspond to the 23-register search space. For example, Loss improves from 62.7\% to 75.5\% AUC and from 9.2\% to 22.9\% TPR@5\%, while Recall improves from 61.7\% to 73.4\% AUC and from 5.6\% to 23.2\% TPR@5\%.
These results support that exact prompt recovery might be unnecessary for auditing. Even when the recovered prompt differs from the true laundering prompt, the synthesized queries can restore candidate-reference separation and make existing detectors effective again. Detailed per-prompt results, including the original and \ourmethod-recovered prompts, are provided in Appendix~\ref{App:specific}. We further test mixed-register laundering prompts in Appendix~\ref{app:mixed}, where \ourmethod~also improves detection performance.

\begin{table}[t]
\centering
\caption{\small Comparison of average performance of unauthorized training data detection across \textbf{different datasets} trained with synthesis using outside register prompts. }
\label{tab:datasets}
\setlength{\tabcolsep}{4pt} % 调整列间距
\renewcommand{\arraystretch}{0.9} % 调整行间距
\footnotesize
\begin{tabular}{lccc|ccc|ccc}
\toprule
\multirow{2}{*}{Method}   & \multicolumn{3}{c}{Wikipedia} & \multicolumn{3}{c}{Hackernews} & \multicolumn{3}{c}{C4} \\
\cmidrule(lr){2-4} \cmidrule(lr){5-7} \cmidrule(lr){8-10}
 & AUC & ASR & TPR@5\% & AUC & ASR & TPR@5\% & AUC & ASR & TPR@5\% \\
\midrule
Recall & 61.7\% & 61.4\% & 5.6\%  & 53.9\% & 56.2\% & 9.9\%  & 52.2\% & 55.0\% & 5.0\% \\
Recall+\ourmethod~  & \textbf{73.4\%} & \textbf{73.3\%} & \textbf{23.2\%} & \textbf{61.6\%} & \textbf{60.3\%} & \textbf{10.7\%} & \textbf{65.2\%} & \textbf{63.0\%} & \textbf{12.0\%} \\
\midrule
Loss   & 62.7\% & 62.6\% & 9.2\%  & 54.2\% & 56.0\% & 6.3\%  & 58.4\% & 59.1\% & 7.2\% \\
Loss+\ourmethod~    & \textbf{75.5\%} & \textbf{75.5\%} & \textbf{22.9\%} & \textbf{62.7\%} & \textbf{59.6\%} & \textbf{8.7\%} & \textbf{67.3\%} & \textbf{64.2\%} & \textbf{13.0\%} \\
\midrule
Min-K  & 64.2\% & 62.5\% & 10.5\% & 53.8\% & 55.9\% & 5.7\%  & 57.6\% & 59.2\% & 6.2\% \\
Min-K+\ourmethod~   & \textbf{73.6\%} & \textbf{73.5\%} & \textbf{22.7\%} & \textbf{61.7\%} & \textbf{60.8\%} & \textbf{8.3\%} & \textbf{66.8\%} & \textbf{64.7\%} & \textbf{11.8\%} \\
\bottomrule
\end{tabular}

\end{table}

\begin{table}[!t]
\centering
\caption{\small Comparison of average performance of unauthorized training data detection across three \textbf{different model architectures} fine-tuned with synthesis generated by outside register prompts. }
% \vspace{-1em}
\label{tab:model}
\setlength{\tabcolsep}{4pt} % 缩小列间距
\renewcommand{\arraystretch}{0.9} % 行距稍微放宽
\footnotesize
\begin{tabular}{lccc|ccc|ccc}
\toprule 
\multirow{2}{*}{Method}   & \multicolumn{3}{c}{Pythia-6.9B} & \multicolumn{3}{c}{Falcon-7B} & \multicolumn{3}{c}{LLama-2-7B} \\
\cmidrule(lr){2-4} \cmidrule(lr){5-7} \cmidrule(lr){8-10}
& AUC & ASR & TPR@5\% & AUC & ASR & TPR@5\% & AUC & ASR & TPR@5\% \\
\midrule
Recall & 61.7\%  & 61.4\%  & 5.6\%  & 62.1\%  & 61.8\%  & 8.2\%  & 61.6\%  & 61.6\%  & 8.2\% \\
Recall+\ourmethod~  & \textbf{73.4\%}  & \textbf{73.3\%}  & \textbf{23.2\%} & \textbf{72.4\%}  & \textbf{69.1\%}  & \textbf{23.0\%} & \textbf{67.5\%}  & \textbf{66.2\%}  & \textbf{12.5\%} \\
\midrule
Loss   & 62.7\%  & 62.6\%  & 9.2\%  & 64.4\%  & 64.4\%  & 11.5\% & 64.5\%  & 64.4\%  & 11.1\% \\
Loss+\ourmethod~    & \textbf{75.5\%}  & \textbf{75.5\%}  & \textbf{22.9\%} & \textbf{71.2\%}  & \textbf{68.0\%}  & \textbf{20.2\%} & \textbf{73.6\%}  & \textbf{70.4\%}  & \textbf{26.9\%} \\
\midrule
Min-K  & 64.2\%  & 62.5\%  & 10.5\%  & 63.3\%  & 62.6\%  & 10.5\% & 62.5\%  & 62.5\%  & 13.9\% \\
Min-K+\ourmethod~   & \textbf{73.6\%}  & \textbf{73.5\%}  & \textbf{22.7\%} & \textbf{70.2\%}  & \textbf{68.0\%}  & \textbf{20.0\%} & \textbf{70.9\%}  & \textbf{68.2\%}  & \textbf{22.9\%} \\
\bottomrule
\end{tabular}
\end{table}
% \textbf{Result analysis across different LLM structures.} 

\subsection{Transferability Results: Robustness under Different Scenarios}

\textbf{Robustness across datasets and target models.} We next evaluate whether \ourmethod~generalizes beyond the default Wikipedia/Pythia setting. Table~\ref{tab:datasets} reports results on three MIMIR subsets under outside-register laundering prompts. \ourmethod~also improves detection across all datasets, indicating that \ourmethod~is not tied to a single data distribution. 
% For example, Recall gains 11.7 AUC points on Wikipedia, 7.7 points on HackerNews, and 13.0 points on C4. Loss and Min-K show similar improvements, indicating that \ourmethod~is not tied to a single data distribution.
Table~\ref{tab:model} evaluates robustness across target model architectures. We fine-tune Pythia-6.9B, Falcon-7B, and Llama-2-7B on laundered MIMIR-Wikipedia data generated from outside-register prompts. \ourmethod~improves all three detector families across all model architectures, indicating that \ourmethod~is not specific to one target architecture.
% For instance, Loss+\ourmethod~improves AUC from 62.7\% to 75.5\% on Pythia-6.9B, from 64.4\% to 71.2\% on Falcon-7B, and from 64.5\% to 73.6\% on Llama-2-7B. These results suggest that \ourmethod~is not specific to one target architecture.

\textbf{Robustness to auxiliary-model mismatch.} A realistic auditor should not know which model was used 
to generate the laundered training data. We therefore test whether 
\ourmethod~remains effective when the auxiliary LLM used by the auditor differs from the laundering model. 
In Table~\ref{tab:llms}, the model provider launders data with 
GPT-4o, while \ourmethod~uses GPT-4o, Claude, or DeepSeek as the auxiliary model. 
The gains remain considerable, indicating that \ourmethod~does not require the auditor's auxiliary model to match the model provider's laundering model.
\begin{wrapfigure}{r}{0.45\textwidth}
    % \vspace{-1em}
    \centering
    \includegraphics[width=1\linewidth]{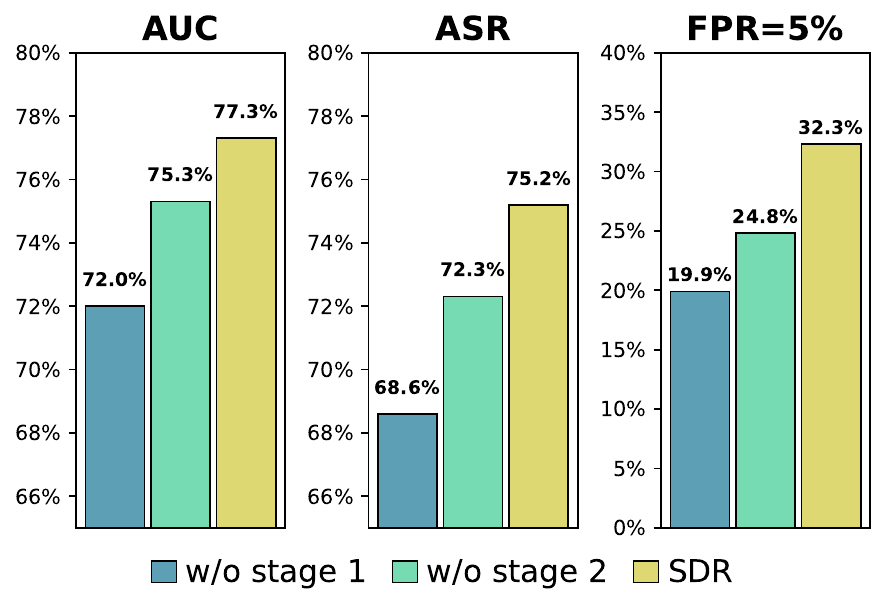}
    % \vspace{-1em}
    \caption{\small Ablation study on the effectiveness of each stage in \ourmethod. 
Results are reported as the average performance of unauthorized training data detectors across different inside prompts. 
Removing the goal identification stage (\textcolor{stage1blue}{w/o stage 1}) 
or the detailed prompt condition inference stage (\textcolor{stage2orange}{w/o stage 2}) 
leads to noticeable degradation, while the full \textcolor{SDRgreen}{\ourmethod~} consistently achieves the best performance. }
    \label{fig:ablation}
    \vspace{-5em}
    \label{fig:0_distillation_concept}
\end{wrapfigure}
% Averaged across detectors, \ourmethod~improves AUC by 13.5 points with GPT-4o, 11.5 points with Claude, and 12.7 points with DeepSeek. This indicates that \ourmethod~does not require the auditor's auxiliary model to match the model provider's laundering model.

We also evaluate stronger transferability settings in Appendix~\ref{app:third_party}. In particular, we test third-party laundering by DeepSeek-v3 and human rewriting using the Polite dataset~\citep{wang2022pay}. \ourmethod~remains generally beneficial in these settings, showing that it can produce detection-useful transformations even when the laundering pipeline differs from the auditor's auxiliary model.

\begin{table}[!t]
\centering
\caption{\small Comparison of the average performance of unauthorized training data detection with \ourmethod~using \textbf{different auxiliary LLMs}.}
\label{tab:llms}
\setlength{\tabcolsep}{4pt}
\renewcommand{\arraystretch}{0.9}
\footnotesize
\begin{tabular}{lccc|ccc|ccc}
\toprule
\multirow{2}{*}{Method}  & \multicolumn{3}{c}{GPT-4o} & \multicolumn{3}{c}{Claude} & \multicolumn{3}{c}{DeepSeek} \\
\cmidrule(lr){2-4} \cmidrule(lr){5-7} \cmidrule(lr){8-10}
 & AUC & ASR & TPR@5\% & AUC & ASR & TPR@5\% & AUC & ASR & TPR@5\% \\
\midrule
Recall & 64.5\% & 63.5\%  & 9.9\% & 66.8\% & 65.9\% & 9.7\% & 65.5\% & 64.8\% & 7.4\% \\
Recall+\ourmethod~  & \textbf{79.7\%} & \textbf{75.0\%}  & \textbf{31.6\%} & \textbf{81.6\%} & \textbf{77.7\%} & \textbf{31.0\%} & \textbf{79.8\%} & \textbf{75.5\%} & \textbf{31.2\%} \\
\midrule
Loss   & 63.4\% & 62.2\%  & 14.3\% & 67.3\% & 65.3\% & 12.1\% & 65.9\% & 64.7\% & 12.2\% \\
Loss+\ourmethod~    & \textbf{80.3\%} & \textbf{75.6\%}  & \textbf{32.6\%} & \textbf{83.4\%} & \textbf{77.8\%} & \textbf{41.9\%} & \textbf{79.0\%} & \textbf{74.7\%} & \textbf{32.8\%} \\
\midrule
Min-K  & 63.7\% & 62.3\%  & 11.0\% & 68.1\% & 65.5\% & 14.5\% & 67.0\% & 64.6\% & 14.2\% \\
Min-K+\ourmethod~   & \textbf{72.0\%} & \textbf{75.1\%}  & \textbf{32.8\%} & \textbf{81.2\%} & \textbf{75.3\%} & \textbf{34.7\%} & \textbf{77.8\%} & \textbf{73.1\%} & \textbf{30.3\%} \\
\bottomrule
\vspace{-3em}
\end{tabular}
\end{table}
\subsection{Ablation Study: Both Goal Identification and Detail Refinement Matter}

\vspace{-0.5em}We conduct ablation studies to evaluate the contribution of the two stages in \ourmethod. Figure~\ref{fig:ablation} compares 
the full method with two variants: removing goal identification (w/o stage 1) and removing detail refinement (w/o stage 2). It can be seen that both stages clearly matter for the performance.
% Removing detail refinement reduces performance across all metrics, showing that a coarse register goal alone is insufficient to approximate the training-time surrogate distribution. In particular, TPR@5\% drops by 7.5 points compared with the full \ourmethod. Removing goal identification causes an even larger degradation, reducing TPR@5\% by 12.5 points. This shows that detail refinement is most effective when initialized from a reliable coarse goal.
Figure~\ref{fig:ablation} also validates the two-stage design: goal identification narrows the search to a plausible register-level direction (with large performance drop after removing stage 1), while detail refinement adjusts the prompt toward a more detection-useful synthesis process. Hyperparameter sensitivity analysis for the iteration budget \(K\), sample sizes \(l\), \(m\), and \(n\) are in Appendix~\ref{app:hyper}.

\vspace{-0.5em}\subsection{Negative Control: \ourmethod~Does Not Fabricate Evidence without Laundering}

\vspace{-0.5em}Because \ourmethod~searches for prompts that maximize detector performance, it is important to verify that it does not create false evidence when the proprietary data was not used for training. We therefore conduct a negative-control experiment where neither \(D_{\mathrm{pro}}\) nor any laundered variant of it appears in the target model's training data. In this setting, \ourmethod~should fail to find a prompt that improves detection (i.e., AUC and ASR should be near 0.5), which is verified by results in Appendix~\ref{app:fpr}.

% The results in Appendix~\ref{app:negative} confirm this behavior. Across ArXiv and Wikipedia, AUC and ASR remain close to 0.5 both before and after \ourmethod, indicating that the prompt search does not by itself create a spurious candidate-reference gap. This sanity check supports the interpretation that \ourmethod's improvements in the main experiments come from restoring laundering-related signals rather than overfitting the detector objective.

\subsection{Additional Stress Tests}

We include several additional evaluations in the appendix. First, Appendix~\ref{app:partial} evaluates partial laundering, where only a subset of \(D_{\mathrm{pro}}\) is laundered and used for training; \ourmethod~still improves all evaluated detectors. Second, Appendix~\ref{app:reverse} compares \ourmethod~with an adapted reverse prompt engineering baseline~\citep{li2024reverse}; \ourmethod~achieves stronger detection gains, highlighting the difference between recovering inference-time prompts and approximating laundering transformations applied before training.

\section{Conclusion}
This paper identifies data laundering as a critical blind spot in post-hoc unauthorized training data detection: when proprietary data is transformed into semantics-preserving surrogates before training, querying the target model with the originals may no longer reveal memorization-based signals. 
We propose \ourmethod, a two-stage method that searches for a detection-useful synthesis prompt rather than recovering the exact laundering transformation. \ourmethod\ first identifies a coarse register-level goal and then refines fine-grained details using target-model feedback, producing training-like queries that allow off-the-shelf detectors to be reused. Experiments across diverse scenarios show that \ourmethod\ consistently improves standard detection methods. 
These results suggest that detection-useful synthesis is a practical direction for auditing unauthorized data use under data laundering. 

%% file: sections/appendix.tex
\newpage
\appendix
\section*{Appendix}
\section{Mathematical Notation}
\label{app:math}
We summarize the key mathematical symbols used throughout the paper.

\begin{table}[ht]
\centering
\renewcommand{\arraystretch}{1.2}
\setlength{\tabcolsep}{8pt}
\footnotesize
\begin{tabular}{cl}
\toprule
Symbol & Description \\
\midrule
$D_{\mathrm{pro}}$ & Proprietary dataset (candidate corpus). \\
% $D_{\mathrm{pro}}'$ & Unknown in-training subset of $D_{\mathrm{pro}}$. \\
$D_{\mathrm{train}}$ & Model provider’s full training dataset. \\
% $D_{\mathrm{lau}}$ & Laundered version of $D_{\mathrm{pro}}'$.\\
$D_{\mathrm{held}}$ & Held-out non-training reference dataset. \\
$M_t$ & Target LLM trained on $D_{\mathrm{train}}$. \\
$M_a$ & Auxiliary LLM used only for synthesis.  \\
$p$ & Current reverse-synthesis prompt refined during iterations. \\
$\mathcal{R}$ & Set of 23 established registers. \\
$r$ & A register in $\mathcal{R}$. \\
$r^\ast$ & Register selected as most closely aligned with the laundering goal. \\
$n, m, l$ & Sample sizes used in constructing templates, scoring, and inference, respectively. \\
$K$ & Maximum number of iterations in the detail refinement stage. \\
$\text{Standard-prompt}_r$ & A canonical prompt that synthesizes data into register $r$. \\
$T$ & True (unknown) laundering transformation. \\
$T_r$ & Opening template extracted for register $r$. \\
$s$ & Original sample. \\
$\hat{s}$ & Synthetic data generated by $M_a$ from an original sample $s$ under prompt $p$. \\
$\tilde{s}$ & Continuation produced by the target model $M_t$ when prompted with $\hat{s}$. \\
$\mathrm{Conf}(r)$ & Continuation confidence of $M_t$ under register $r$. \\
$\mathcal{C}$ & Candidate set of top-$k$ registers with highest confidence scores. \\
$\mathrm{Syn}_r$ & Dataset synthesized into register $r$ by $M_a$. \\
$\mathrm{Perf}_r$ & Unauthorized training data detection performance of $\mathrm{Syn}_r$ on $M_t$. \\
$\mathrm{Syn}_p$ &  Dataset synthesized into register $p$ by $M_a$. \\
$\mathrm{Perf}_p$ &  Unauthorized training data detection performance of $\mathrm{Syn}_p$ on $M_t$.\\
\bottomrule
\end{tabular}
\end{table}

\section{Register Taxonomy}
\label{app:subregister}

The register taxonomy proposed by \citet{henriksson2024automatic} defines 23 sub-registers, ranging from narrative forms (e.g., news and sports reports, blogs) and informational texts (e.g., encyclopedia entries, research articles, legal documents) to opinion pieces, persuasive descriptions, interactive discussions (e.g., FAQs, interviews), instructional texts (e.g., recipes), spoken and lyrical registers (details are shown in Table~\ref{tab:rigster}).
This taxonomy provides a broad linguistically motivated coverage of common open-domain registers. 
In our work, it is used as a systematic and scalable framework that reduces unbounded stylistic variation into a bounded set of functional categories.

\begin{table}[htbp]
\centering
\caption{Sub-registers and abbreviations.}
\label{tab:rigster}
\vspace{4pt}\footnotesize
\begin{tabular}{llcll}
\toprule
Name & Abbr. & & Name & Abbr. \\
\midrule
Lyrical & ly & & Encyclopedia article & en \\
Spoken & sp & & Research article & ra \\
Interview & it & & Description of a thing or person & dtp \\
Interactive discussion & id & & FAQ & fi \\
Narrative & na & & Legal terms \& conditions & lt \\
News report & ne & & Opinion & op \\
Sports report & sr & & Review & rv \\
Narrative blog & nb & & Opinion blog & ob \\
How-to or instructions & hi & & Denominational religious blog or sermon & rs \\
Recipe & re & & Description with intent to sell & ds \\
Informational persuasion & ip & & Informational description & in \\
News \& opinion blog or editorial & ed & & & \\
\bottomrule
\end{tabular}
\end{table}

\section{The Objective and Threat Model}
\label{app:threat}
\textbf{Threat model.}
Our setting follows a black-box auditing formulation used in post-hoc unauthorized data detection~\citep{xie2024recall, carlini2022membership} where:
1. The model provider trains the target LLM $M_t$ on an unknown training set $D_{\mathrm{train}}$ that may include laundered variants of some samples in a proprietary dataset $D_{\mathrm{pro}}$.
2. The data rights holder owns a proprietary dataset $D_{\mathrm{pro}}$ and has a held-out reference set $D_{\mathrm{held}}$ guaranteed not to appear in $D_{\mathrm{train}}$. The rights holder only has access to (i) $D_{\mathrm{pro}}$, (ii) $D_{\mathrm{held}}$, and (iii) black-box query access to $M_t$ (e.g., API). It does \textbf{not} have access to $D_{\mathrm{train}}$, nor to the (potentially unknown) laundering pipeline.
3. An auxiliary LLM $M_a$ is any model that the rights holder can query to synthesize surrogate samples. 
Then, we aim to help rights holders detect whether $M_t$ used $D_{\mathrm{pro}}$ or laundered variants of $D_{\mathrm{pro}}$ for its training.
We additionally summarize all symbols and their roles in Table~\ref{tab:notation}.

\begin{table}[!t]
\caption{Notation list for the threat model and if they are visible to rights holder.}
\centering
\begingroup
\setlength{\tabcolsep}{6pt} % 控制列宽（数字越小越窄）% 交替行蓝色背景
\vspace{4pt}\footnotesize
\begin{adjustbox}{max width=\linewidth} % 自动缩放到页宽
\begin{tabular}{lll}
\toprule
Symbol & Description \& Role & Visible to Rights Holder? \\
\midrule
$M_t$ & Target LLM trained on $D_{\mathrm{train}}$ & No \\
$M_a$ & Auxiliary LLM used only for the prompt reversion. & Yes \\
$T$ & True (unknown) laundering transformation. & No \\
$p$ & Reverse-synthesis prompt inferred by \ourmethod. & Yes \\
$D_{\mathrm{pro}}$ & Proprietary dataset (candidate corpus). & Yes \\
% $D_{\mathrm{pro}}'$ & Unknown in-training subset of $D_{\mathrm{pro}}$ & No \\
$D_{\mathrm{train}}$ & Model provider’s full training dataset. & No \\
% $D_{\mathrm{lau}}$ & Laundered version of $D_{\mathrm{pro}}'$ & No \\
$D_{\mathrm{held}}$ & Held-out non-training reference dataset. & Yes \\
$s$ & Original sample. & Yes \\
$\hat{s}$ & Synthetic surrogate rewrite. & Yes \\
$\mathrm{Syn}_p$ & Dataset synthesized into register $p$ by $M_a$. & Yes \\
$\mathrm{Perf}_p$ & Unauthorized training data detection performance of $\mathrm{Syn}_p$ on $M_t$. & Yes \\
\bottomrule
\end{tabular}
\end{adjustbox}
\endgroup
\label{tab:notation}
\end{table}

\textbf{What \ourmethod~optimizes.}
\ourmethod~seeks to ``optimize'' a prompt $p$ so that samples in $D_{\mathrm{pro}}$ are mapped closer to the true laundered variants used in training $M_t$, thereby restoring the effectiveness of standard unauthorized training data detectors.
Firstly, We use $M_a$ to synthesize $\text{Syn}_p(D_{\rm pro})$ and $\text{Syn}_p(D_{\rm held})$. And then runs an off‑the‑shelf detector (Recall, Min‑K, etc.) on $M_t$ using these two sets, exactly as in standard post‑hoc unauthorized data detection, but using the synthesized variants. Finally, \ourmethod~takes the detector’s output scalar score (e.g., Recall’s two‑sample score) as $\text{Perf}_p$, and \ourmethod~seeks to optimize a prompt $p$ that maximizes $\text{Perf}_p$.

\textbf{Detailed definition of $\mathrm{Perf}_p$.} 
Given a fixed detector (e.g., Recall, Min‑K), a prompt $p$ induces synthetic surrogates ${\rm Syn}_p (D_{\rm pro})$ and ${\rm Syn}_p (D_{\rm held})$ via $M_a$. We then run the detector on $M_t$ precisely as in prior work, but using these surrogates as candidate vs. reference sets, and record a scalar performance measure ${\rm Perf_p}$. Algorithms~\ref{alg:directive} and~\ref{alg:condition} describe our pipeline that searches over the prompt space to maximize this scalar.
${\rm Perf_{p}}$ distinguishes ${\rm Syn}_p (D_{\rm pro})$ from ${\rm Syn}_p (D_{\rm held})$, implying that $M_t$ was indeed trained on a laundered version of $D_{\rm pro}$, and $p$ can recover the unknown laundering process used by the model provider.
Otherwise, it implies that the current prompt $p$ cannot recover the laundered training data, and we need to continue searching for more plausible ones.
Eventually, if no such performance-improving prompt can be found, we consider that no laundering of $D_{\rm pro}$ was used in training $M_t$.

\vspace{-1em}
\section{Opening Template}
\label{app:template}
Table~\ref{tab:register-templates} presents the representative opening templates $T_r$ that we derived for each register. These templates were obtained by synthesizing a small subset of samples into the corresponding register $r$, after which a large language model was used to extract a generalized first-sentence structure from these synthesized samples.
As shown in the table, each register exhibits distinct stylistic cues in its openings. For example, lyrical texts often begin with abstract imagery, interviews with a direct address from the interviewer, and storytelling narratives with a scene-setting phrase such as “Once upon a time.”

\begin{table}[H]
\centering
\caption{Registers and their corresponding opening templates.}
\label{tab:register-templates}
\footnotesize
\begin{tabularx}{\textwidth}{lX}
\toprule
Register & Opening Template \\
\midrule
Lyrical & In the heart of [abstract domain], a tale unfolds, where [abstract concept], [abstract detail], [abstract entity], [abstract action]. \\

Spoken style & So, let’s talk about [TOPIC]. \\

Interview & \textbf{Interviewer:} Thank you for joining us, [Person/Expert Title]. Can you tell us about [Subject/Topic]? \\

Interactive discussion & \textbf{[Participant 1]:} So, have you guys heard about [Topic/Subject]? I recently came across some interesting information about it. \\

Storytelling narrative & Once upon a time, in a [adjective] [type of place] called [place name], there lived a [adjective] [type of character] named [character name]. \\

News report & \textbf{[Event/Topic]: [Description/Significance] [Location/Context]} – [Details about the subject, including noteworthy contributions, roles, or milestones]. \\

Sports report & In a thrilling [event/display/action], [subject/actor] has [verb] [description/impact] in [field/area/genre]. \\

Narrative blog post & In the context of [broad category or field], [subject or specific work] has made a significant impact, often leading to [general observation or effect]. \\

Step-by-step guide (How-to) & \textbf{Step-by-Step Guide to Understanding [Subject]} — Step 1: [Initial focus or background]. Learn that [Subject Description]. \\

Recipe & \textbf{Recipe for [General Concept]: [Specific Edition/Style]} — \textit{Ingredients:} [Variable 1], [Variable 2], [Variable 3]\ldots \\

Encyclopedia article & \textbf{[Subject]} is a [type/category] that [provides a description or function], [additional information if applicable]. \\

Research article & This article explores the significance of [subject or topic], a [description or classification], characterized by [notable features or contributions]. \\

Description of a thing or person & Introducing [Subject/Entity], a [descriptor] [type/category] [context/detail] renowned for its [property/characteristic]. \\

FAQ & \textbf{What is [Subject]?} — [Subject] is a [general category or description] [specific type or detail] [additional information]. \\

Legal terms \& conditions & \textbf{Terms and Conditions Regarding [Subject/Theme]}. \\

Opinion & In my view, [Subject/Entity] represents [significance/impact/legacy] in [field/area], and its influence on [audience/community/context] cannot be overstated. \\

Review & [Subject] is a [descriptor] that [verb phrase] [contextual information]. \\

Opinion blog (editorial) & When we think of [general category or field], [a notable example or subject] often comes to mind. \\

Denominational religious sermon & Beloved congregation, today we gather to reflect upon [individual/concept] that illuminates our lives and encourages us to contemplate our shared journey. \\

Description with intent to sell & Introducing [Subject]: a [descriptor] [product/service] designed for [use case]; discover how it [benefit/outcome] for [target user]. \\

Informational persuasion & In the context of [domain or field], few [types/categories] resonate as profoundly within [subfields] as [specific work/name/entity]. \\

Informational description & [Entity/Subject] is a [description] in the field of [broader category], specifically within [subcategory/locale]. \\

News \& opinion blog or editorial & When we think of [general category or field], [notable subject] often comes to mind — situating today’s discussion of [topic] within [context]. \\
\bottomrule
\end{tabularx}
\end{table}

\section{Benchmark Details}
\label{app:mimir}
We evaluate our method on the MIMIR benchmark, introduced by \citet{deng2023investigating} as a standardized evaluation framework for detecting unauthorized training data in large language models. 
The benchmark consists of several domain-specific subsets, including Wikipedia articles, news articles, and general-domain web texts (i.e., C4).
For each domain-specific subset, MIMIR provides both training and non-training samples. Training samples are drawn from data used during model training, whereas non-training samples are collected from sources not included in the training corpus of target models (including Pythia, Falcon, and LLama-2). 
In all experiments presented in this paper, we exclusively use samples from the non-training portion of MIMIR. 
Specifically, a subset of non-training samples is randomly selected and transformed into different registers to construct the laundering dataset used for prompt reversion. Another disjoint subset of non-training samples is reserved as held-out data for evaluation. 
The MIMIR dataset is publicly available at
\url{https://huggingface.co/datasets/iamgroot42/mimir}.

\section{Experimental Setup for Laundering Reversal Evaluation}
\label{sec:detail implement}
\subsection{Evaluation Metrics}
\label{sec:metrics}
Following \citet{carlini2022membership}, we adopt three complementary metrics to evaluate the detection of unauthorized training data.

\textbf{Area Under the ROC Curve (AUC).}
AUC measures the overall discriminative power of the detection, independent of any specific threshold. It reflects how well an unauthorized training data detection method can separate training data from unseen data on average. It may overstate effectiveness because it also includes high-false-positive regions that are less relevant in practice.

\textbf{Attack Success Rate (ASR).}
ASR measures the fraction of correctly identified training data under a single decision threshold that maximizes balanced accuracy across training data and unseen data. Unlike AUC, ASR reflects the practical effectiveness of a detector when deployed, as real-world unauthorized training data detectors typically operate at a single fixed threshold.

\textbf{True Positive Rate at 5\% False Positive Rate (TPR@5\%).}
This metric evaluates a detector's ability to identify training data while maintaining a strict false-positive constraint. 
Prior work highlights that low false-positive regimes are most meaningful for privacy evaluation, since even a small number of incorrect training-data decisions can undermine the credibility of the detection. 
TPR@5\% therefore provides a high-precision view of detection success.

\subsection{Experimental Setup}
We evaluate the effectiveness of our approach by examining whether \ourmethod-synthesized data enhances the performance of existing unauthorized training data detection methods against LLMs trained on synthesized data. 
Specifically, we first sample 200 samples from the dataset that have not been seen by the {\em target} LLM. 
These 200 samples are synthesized into a new version using an unknown prompt that simulates the data-laundering process applied by the model provider.
We then randomly select half of these synthesized samples as training data to fine-tune the LLM (training details are provided in Table~\ref{tab:training}), while the remaining half serves as held-out data. 
Subsequently, \ourmethod~is employed to infer a transformation prompt that restores the synthesis process and recovers the synthesized samples.
Using this inferred prompt, we rewrite the original 200 samples. 
We apply baseline unauthorized training data detectors to both the original 200 samples and the 200 samples rewritten with the inferred prompt, and compare their detection performance to assess whether the inferred prompt improves detection effectiveness.

\begin{table}[ht]
\caption{Hyperparameter settings for LoRA fine-tuning.}
\centering
\vspace{4pt}\footnotesize
\begin{tabular}{c|c|c|c|c|c|c}
\toprule
Lora\_alpha & r & epochs & lr & \makecell{gradient\\accumulation\\steps} & weight\_decay & logging\_strategy \\
\midrule
32 & 8 & 30 & 0.0004 & 25 & 0.1 & ``epoch'' \\
\bottomrule
\end{tabular}
\label{tab:training}
\end{table}

\section{Laundering Prompts Used in Experiments}
\label{app:prompts}
Table~\ref{tab:inside} lists the complete set of inside-register prompts corresponding to the 23 predefined registers in our framework. Each prompt is generated by GPT-5 using the instruction: “Generate a rewriting prompt that transfers the text into [REGISTER].” Here, [REGISTER] denotes one of the 23 registers.
Table~\ref{tab:outside} lists the complete set of outside-register prompts; none align with any of the 23 predefined registers. These prompts are generated by GPT-5 using the instruction: “You are a prompt generator. Generate rewriting prompts that transform the original content into broadly different domains (e.g., Facebook post, academic article, children’s story, legal document). Make the prompts as mutually distinct in domain as possible, and ensure each induces substantial changes to the original text rather than merely surface-level edits.”

\begin{table}[htbp]
\centering
\caption{Inside-register prompts.}
\label{tab:inside}
\vspace{4pt}\footnotesize
\begin{tabularx}{\textwidth}{rX}
\toprule
ID & Laundering Prompts \\
\midrule
1  & Rewrite the text in a lyrical style, ensuring the imagery is vivid, the rhythm flows naturally. \\
2  & Rewrite the text in a spoken style, making it sound natural and conversational, and ensure the tone feels engaging and easy to follow for a live audience. \\
3  & Rewrite the text in the form of an interview, ensuring the questions flow naturally and the answers provide clear, engaging explanations for the audience. \\
4  & Rewrite the text as an interactive discussion between two or more participants, ensuring the conversation flows logically, with each speaker’s tone and style clearly distinguishable. \\
5  & Rewrite the text as a storytelling narrative. The story should flow naturally, use simple and engaging language, and be easy for all kinds of listeners to follow. \\
6  & Rewrite the text in the style of a news report, ensuring the information is presented objectively and concisely. \\
7  & Rewrite the text as a sports report, ensuring the action is described with dynamic, energetic language that conveys the pace, tension, and excitement of the event. \\
8  & Rewrite the text as a narrative blog post, organized into clear sections with subheadings. Use a tone that is engaging and reflective, blending storytelling with explanation. \\
9  & Rewrite the text as a step-by-step instructional guide. Break the content into numbered steps, with each step beginning with a clear imperative verb. \\
10  & Rewrite the text as a recipe, introduce the information as sequential steps. \\
11 & Rewrite the text to persuade the reader through factual information, making sure to include at least three specific data points or statistics to support the argument. \\
12 & Rewrite the text as a sales description, and be sure to include a clear call-to-action at the end. \\
13 & Rewrite the text in the style of an editorial, making sure to include a clear stance or opinion and a concluding paragraph that calls for action or reflection. \\
14 & Rewrite the text as an informational description, ensuring the tone is neutral and objective, and include at least one definition or clarification to help the reader better understand the subject. \\
15 & Rewrite the text in the style of an encyclopedia entry, maintaining a neutral, authoritative tone, and include at least one date, fact, or reference to give it the appearance of being sourced. \\
16 & Rewrite the text as an academic research article, structured with sections such as Abstract, Introduction, Method, Results, and Conclusion, and include at least one in-text citation (invented if necessary) to simulate scholarly referencing. \\
17 & Rewrite the text as a descriptive profile of a specific thing or person, using vivid details and attributes (appearance, characteristics, or context) and ending with a short summary sentence that highlights its significance. \\
18 & Rewrite the text in the form of a Frequently Asked Questions (FAQ) section, making sure to include at least three question--answer pairs, with the questions phrased from the perspective of a curious reader. \\
19 & Rewrite the text as legal terms and conditions, using formal legal language, and ensure at least one numbered clause is included for clarity. \\
20 & Rewrite the text as a personal opinion piece, written in the first person, making sure to clearly express a stance and support it with at least one reason or example. \\
21 & Rewrite the text as a review, giving it a clear positive or negative stance, and include at least one specific detail or example to justify the evaluation. \\
22 & Rewrite the text as an opinion blog post, written in a conversational and persuasive tone, and include at least one personal anecdote or illustrative example to strengthen the argument. \\
23 & Rewrite the text as a denominational religious sermon, using a reverent and exhortative tone, and include at least one scriptural quotation or moral teaching to guide the audience toward reflection or action. \\
\bottomrule
\end{tabularx}
\end{table}

\begin{table}[H]
\centering
\caption{Outside-register prompts.}
\label{tab:outside}
\vspace{4pt}\footnotesize
\begin{tabularx}{\textwidth}{rX}
\toprule
ID & Laundering Prompts \\
\midrule
1  & Rewrite the following content as slide presentation bullet points. Focus on summarizing the key arguments and findings clearly and concisely. Use concise phrases that highlight core points. \\
2  & Rewrite the following text in the style of a Facebook post. Sharing interesting information with followers. You may add light commentary, questions to the audience, or casual phrasing, but keep it natural and human-like. Avoid using emojis, hashtags, or overly dramatic expressions. \\
3  & Adapt the text into a poetic form with vivid metaphors, rhythmic structure, and emotionally evocative language. \\
4  & Convert the content into a tutorial-style explanation for beginners, using step-by-step instructions, simple analogies, and common misunderstandings. \\
5  & Rewrite the text as a formal business email, ensuring clarity, professionalism, and a polite tone. \\
6  & Rewrite the passage as a scientific abstract, including Background, Methods, Results, and Conclusions. Invent at least two numerical values (percentages, sample sizes, or statistical outcomes) to support claims. \\
7  & Rewrite the text as a product description for an e-commerce website, highlighting key features, benefits, and use cases in a persuasive manner. \\
8  & Rewrite the text as a blog post, incorporating vivid descriptions of locations, cultural insights, and personal experiences to engage readers. \\
9  & Rewrite the text as a classroom lecture transcript, with explanations, rhetorical questions, and occasional student interaction. \\
10 & Rewrite the text with stronger transitions between sentences and paragraphs, ensuring smoother reading without adding new information. \\
\bottomrule
\end{tabularx}
\end{table}

\section{The Specific Results for Each Laundering Prompt}
\label{App:specific}
Tables~\ref{tab:inside_1} and \ref{tab:inside_2} present the detailed average AUC performance across five unauthorized training data detection methods for each inside-register laundering prompt. Tables~\ref{tab:outside_combined} present the details for each outside-register laundering prompt. 
The ``Original Prompt" column denotes the true laundering prompt. The “Reversed Prompt” column denotes the best prompt recovered by \ourmethod. The “Orig.” column reports the average AUC across detection methods using the original data, whereas the “\ourmethod” column reports the results using \ourmethod-reversed data.

\begin{table}[H]
\centering
\caption{Inside-register prompts and corresponding reversed prompts with Orig and \ourmethod~results.}
\vspace{-1em}
\label{tab:inside_1}
\vspace{4pt}\footnotesize
\begin{tabularx}{\textwidth}{rX X cc}
\toprule
ID & Original Prompt & Reversed Prompt (\ourmethod) & Orig (AUC) & \ourmethod~(AUC) \\
\midrule
1  & Rewrite the text in a lyrical style, ensuring the imagery is vivid, the rhythm flows naturally. 
   & Rewrite the text in a lyrical style, enhancing the poetic rhythm and imagery while capturing the essence and emotional depth of the original content. 
   & 0.540 & 0.692 \\
2  & Rewrite the text in a spoken style, making it sound natural and conversational, and ensure the tone feels engaging and easy to follow for a live audience. 
   & Rewrite the text to sound natural and conversational, using everyday language and personal anecdotes to create an engaging and friendly atmosphere for the listener. 
   & 0.702 & 0.894 \\
3  & Rewrite the text in the form of an interview, ensuring the questions flow naturally and the answers provide clear, engaging explanations for the audience. 
   & Rewrite the text in the form of an interview, ensuring a clear and engaging dialogue that accurately conveys the information while maintaining a conversational tone and eliciting detailed responses. 
   & 0.713 & 0.823 \\
4  & Rewrite the text as an interactive discussion between two or more participants, ensuring the conversation flows logically, with each speaker’s tone and style clearly distinguishable. 
   & Rewrite the text as an interactive discussion between two or more participants, ensuring a natural flow of dialogue that incorporates factual information and engages with the topic through building on each other's comments. 
   & 0.650 & 0.770 \\
5  & Rewrite the text as a storytelling narrative. The story should flow naturally, use simple and engaging language, and be easy for all kinds of listeners to follow. 
   & Rewrite the text in the form of a Frequently Asked Questions section, transforming the information into clear and concise questions and answers that emphasize key details and engage the reader effectively. 
   & 0.646 & 0.735 \\
6  & Rewrite the text in the style of a news report, ensuring the information is presented objectively and concisely. 
   & Rewrite the text into a Frequently Asked Questions section, organizing the information into clear and concise questions and answers while highlighting key details and maintaining clarity and readability.
   & 0.793 & 0.904 \\
7  & Rewrite the text as a sports report, ensuring the action is described with dynamic, energetic language that conveys the pace, tension, and excitement of the event. 
   & Rewrite the text in the style of an engaging editorial, enhancing the narrative through vivid language, emotional depth, and a focus on the significance of the subject matter. 
   & 0.655 & 0.713 \\
8  & Rewrite the text as a narrative blog post, organized into clear sections with subheadings. Use a tone that is engaging and reflective, blending storytelling with explanation. 
   & Rewrite the text in the form of a Frequently Asked Questions section, focusing on clearly structured questions and answers that highlight key aspects, contributions, and significance of the subject matter in a conversational tone. 
   & 0.667 & 0.748 \\
9  & Rewrite the text as a step-by-step instructional guide. Break the content into numbered steps, with each step beginning with a clear imperative verb. 
   & Rewrite the text as a step-by-step instructional guide, breaking down the information into clear, organized steps that highlight key concepts, details, and relevant insights for enhanced understanding. 
   & 0.778 & 0.841 \\
10 & Rewrite the text as a recipe, introduce the information as sequential steps. 
   & Rewrite the text to persuasively present factual information, emphasizing key aspects and structuring the content clearly to enhance engagement and clarity. 
   & 0.700 & 0.724 \\
11 & Rewrite the text to persuade the reader through factual information, making sure to include at least three specific data points or statistics to support the argument. 
   & Rewrite the text in the style of an encyclopedia entry, focusing on enhancing structural clarity, coherence, and technical detail by organizing the information into distinct sections and emphasizing historical significance and key contributions. 
   & 0.633 & 0.633 \\
12 & Rewrite the text as a sales description, and be sure to include a clear call-to-action at the end. 
   & Rewrite the text as a Frequently Asked Questions section, transforming the original content into a clear and engaging question-and-answer format that effectively highlights key elements, significance, and context for the reader. 
   & 0.622 & 0.725 \\
\bottomrule
\end{tabularx}
\end{table}

\begin{table}[H]
\centering
\caption{Inside-register prompts and corresponding reversed prompts with Orig and \ourmethod~results (continues).}
\label{tab:inside_2}
\vspace{4pt}\footnotesize
\begin{tabularx}{\textwidth}{rX X cc}
\toprule
ID & Original Prompt & Reversed Prompt (\ourmethod) & Orig (AUC) & \ourmethod~(AUC) \\
\midrule
13 & Rewrite the text in the style of an editorial, making sure to include a clear stance or opinion and a concluding paragraph that calls for action or reflection. 
   & Rewrite the text as a personal opinion piece, emphasizing reflective commentary and personal insights while exploring the broader societal implications and significance of the subject matter. 
   & 0.594 & 0.657 \\
14 & Rewrite the text as an informational description, ensuring the tone is neutral and objective, and include at least one definition or clarification to help the reader better understand the subject. 
   & Rewrite the text as a step-by-step instructional guide, organizing the content into clear, numbered sections that effectively communicate essential information about the subject. 
   & 0.594 & 0.795 \\
15 & Rewrite the text in the style of an encyclopedia entry, maintaining a neutral, authoritative tone, and include at least one date, fact, or reference to give it the appearance of being sourced. 
   & Rewrite the text as an informational description, focusing on presenting a clear, structured overview of the subject's key facts, achievements, and background while maintaining concise and objective language. 
   & 0.792 & 0.923 \\
16 & Rewrite the text as an academic research article, structured with sections such as Abstract, Introduction, Method, Results, and Conclusion, and include at least one in-text citation (invented if necessary) to simulate scholarly referencing. 
   & Rewrite the text in the form of an interview, transforming the original content into a conversational dialogue that incorporates engaging questions and responses while maintaining clarity and coherence. 
   & 0.525 & 0.546 \\
17 & Rewrite the text as a descriptive profile of a specific thing or person, using vivid details and attributes (appearance, characteristics, or context) and ending with a short summary sentence that highlights its significance. 
   & Rewrite the text as a sales description, transforming it into an engaging narrative that highlights the subject's achievements, legacy, and emotional impact to captivate and appeal to potential audiences. 
   & 0.661 & 0.765 \\
18 & Rewrite the text in the form of a Frequently Asked Questions (FAQ) section, making sure to include at least three question--answer pairs, with the questions phrased from the perspective of a curious reader. 
   & Rewrite the text in the form of an interview, transforming factual information into a conversational question-and-answer format that captures personal insights, key themes, and details from the original content. 
   & 0.640 & 0.792 \\
19 & Rewrite the text as legal terms and conditions, using formal legal language, and ensure at least one numbered clause is included for clarity. 
   & Rewrite the text into a Frequently Asked Questions section by converting the content into clear questions and answers, ensuring clarity, conciseness, accuracy, and structured organization of information. 
   & 0.731 & 0.833 \\
20 & Rewrite the text as a personal opinion piece, written in the first person, making sure to clearly express a stance and support it with at least one reason or example. 
   & Rewrite the text as a review, focusing on summarizing key aspects and implications while maintaining an engaging narrative style that connects with the reader. 
   & 0.575 & 0.613 \\
21 & Rewrite the text as a review, giving it a clear positive or negative stance, and include at least one specific detail or example to justify the evaluation. 
   & Rewrite the text as a review, emphasizing the subject's significance, key achievements, and connections to broader themes or contexts, while maintaining a consistent tone and providing a balanced evaluation. 
   & 0.593 & 0.698 \\
22 & Rewrite the text as an opinion blog post, written in a conversational and persuasive tone, and include at least one personal anecdote or illustrative example to strengthen the argument. 
   & Rewrite the text as a conversational interview, focusing on transforming factual content into dialogue by incorporating questions, responses, and personal insights while maintaining the original essence. 
   & 0.577 & 0.630 \\
23 & Rewrite the text as a denominational religious sermon, using a reverent and exhortative tone, and include at least one scriptural quotation or moral teaching to guide the audience toward reflection or action. 
   & Rewrite the text as a denominational religious sermon, transforming the narrative into an inspirational message that emphasizes spiritual themes, fosters community, and resonates with the congregation's values. 
   & 0.541 & 0.627 \\
\bottomrule
\end{tabularx}
\end{table}

\begin{table}[H]
\centering
\caption{Outside-register prompts and corresponding reversed prompts with Orig and \ourmethod~results.}
\label{tab:outside_combined}
\vspace{4pt}\footnotesize
\begin{tabularx}{\textwidth}{rX X cc}
\toprule
ID & Original Prompt & Reversed Prompt (\ourmethod) & Orig & \ourmethod~\\
\midrule
1  & Rewrite the following content as slide presentation bullet points. Focus on summarizing the key arguments and findings clearly and concisely. Use concise phrases that highlight core points. 
   & Rewrite the text as a step-by-step instructional guide, organizing the information into clear sections and ensuring each step provides concise, relevant details on the specified topic. 
   & 0.730 & 0.799 \\
2  & Rewrite the following text in the style of a Facebook post. Sharing interesting information with followers. You may add light commentary, questions to the audience, or casual phrasing, but keep it natural and human-like. Avoid using emojis, hashtags, or overly dramatic expressions. 
   & Rewrite the text as a Frequently Asked Questions section, transforming the information into an engaging question-and-answer format that encourages reader interaction and maintains a conversational tone. 
   & 0.644 & 0.730 \\
3  & Adapt the text into a poetic form with vivid metaphors, rhythmic structure, and emotionally evocative language. 
   & Rewrite the text in a lyrical style that transforms factual content into an evocative narrative, using vivid imagery, poetic devices, and rhythmic flow to highlight emotional resonance and thematic cohesion. 
   & 0.538 & 0.674 \\
4  & Convert the content into a tutorial-style explanation for beginners, using step-by-step instructions, simple analogies, and common misunderstandings. 
   & Rewrite the text as a step-by-step instructional guide, ensuring clear and concise steps that effectively outline key aspects and concepts while maintaining an engaging tone and logical flow throughout. 
   & 0.616 & 0.679 \\
5  & Rewrite the text as a formal business email, ensuring clarity, professionalism, and a polite tone. 
   & Rewrite the text in the style of an encyclopedia entry, emphasizing clear and concise organization, formal language, and distinct sections that present factual information and key points effectively. 
   & 0.795 & 0.884 \\
6  & Rewrite the passage as a scientific abstract, including Background, Methods, Results, and Conclusions. Invent at least two numerical values (percentages, sample sizes, or statistical outcomes) to support claims. 
   & Rewrite the text as a Frequently Asked Questions section, transforming the original content into clear, concise questions and answers that emphasize key themes, significant information, and factual accuracy. 
   & 0.600 & 0.678 \\
7  & Rewrite the text as a product description for an e-commerce website, highlighting key features, benefits, and use cases in a persuasive manner. 
   & Rewrite the text as a dialogue in an interview format, emphasizing key details and insights while maintaining clarity and engagement through a question-and-answer structure. 
   & 0.603 & 0.673 \\
8  & Rewrite the text as a blog post, incorporating vivid descriptions of locations, cultural insights, and personal experiences to engage readers. 
   & Rewrite the text as a sales description that emphasizes unique aspects and engaging narratives, highlighting significance and emotional appeal to captivate the audience. 
   & 0.626 & 0.728 \\
9  & Rewrite the text as a classroom lecture transcript, with explanations, rhetorical questions, and occasional student interaction. 
   & Rewrite the text in the form of an interview, transforming the information into a natural dialogue that incorporates questions and answers while preserving the original content's key details and themes. 
   & 0.667 & 0.764 \\
10 & Rewrite the text with stronger transitions between sentences and paragraphs, ensuring smoother reading without adding new information. 
   & Rewrite the text in the style of an editorial, focusing on enhancing the narrative through emotional engagement, historical significance, and the subject’s impact, while highlighting community involvement and contemporary relevance. 
   & 0.570 & 0.646 \\
\bottomrule
\end{tabularx}
\end{table}

\section{Negative-control experiment}
\label{app:fpr}

In the negative-control experiment setup, where neither $D_\text{pro}$ nor its laundered variants were used \ourmethod. The result in Table~\ref{tab:fpr} shows that \ourmethod~fails to find any prompt that improves detection: across all detectors, both AUC and ASR remain close to 0.5 on ArXiv and Wikipedia. This demonstrates that \ourmethod~does not spuriously create evidence of data misuse when $D_{\rm pro}$ was not used for training.

\begin{table}[ht]

\caption{Negative-control experiment on \textbf{Pythia-6.9B}. 
When the target model is \emph{not} trained on any laundered variants of the proprietary data, 
\ourmethod~does not find a prompt to improve unauthorized training-data detection. 
Across all detectors and both datasets (ArXiv and Wikipedia), the AUC and ASR scores before and after \ourmethod~remain close to 0.5, 
indicating that \ourmethod~does not fabricate evidence of misuse in the absence of laundered training data.}
\centering
\label{tab:fpr}
\vspace{4pt}\footnotesize
\begin{tabular}{lcc|cc}
\toprule
\multirow{2}{*}{Method} 
& \multicolumn{2}{c}{ArXiv} 
& \multicolumn{2}{c}{Wikipedia} \\
\cmidrule(lr){2-3} \cmidrule(lr){4-5}
& AUC & ASR
& AUC & ASR \\
\midrule
Recall            & 0.516 & 0.532 & 0.528 & 0.545 \\
Recall + \ourmethod      & 0.488 & 0.525 & 0.473 & 0.540 \\
\midrule
Loss              & 0.486 & 0.522 & 0.471 & 0.510 \\
Loss + \ourmethod       & 0.490 & 0.535 & 0.465 & 0.515 \\
\midrule
Min-K             & 0.536 & 0.510 & 0.486 & 0.515 \\
Min-K + \ourmethod      & 0.499 & 0.535 & 0.440 & 0.525 \\
\midrule
Min-K++           & 0.476 & 0.521 & 0.483 & 0.535 \\
Min-K++ + \ourmethod     & 0.483 & 0.520 & 0.446 & 0.515 \\
\bottomrule
\end{tabular}

\label{tab:negative-control}
\end{table}

\section{\ourmethod~Transferability to Third-Party Laundering Pipelines}
\label{app:third_party}
\textbf{Alternative LLM Launderer with DeepSeek.}
We fine-tune the target LLM exclusively on data laundered by DeepSeek-v3~\citep{liu2024deepseek}, while \ourmethod~continues to use GPT-4o as the auxiliary model $M_a$ for prompt search.
We apply the first five inside-register prompts (Table~\ref{tab:inside}) as laundering prompts and perform laundering using DeepSeek-v3.
As shown in Table~\ref{tab:deepseek-transfer}, \ourmethod~consistently improves detection performance across all four off-the-shelf detectors.
For example, Loss improves from AUC/ASR of $0.65/0.64$ to $0.81/0.76$, and Min-K improves from $0.67/0.65$ to $0.78/0.73$.
This demonstrates that \ourmethod~can recover useful transformations even when the synthesized model used in the laundering pipeline is completely different from GPT-4o.

\begin{table}[ht]
\caption{\ourmethod~transferability on DeepSeek-v3 laundering pipelines for the target LLM trained exclusively on DeepSeek-laundered data. \ourmethod~uses GPT-4o as $M_a$.}
\centering
\vspace{4pt}\footnotesize
\begin{tabular}{lcc}
\toprule
Method & AUC & ASR \\
\midrule
Recall & 0.641 & 0.630 \\
Recall + \ourmethod & \textbf{0.767} & \textbf{0.718} \\
\midrule
Loss & 0.650 & 0.635 \\
Loss + \ourmethod & \textbf{0.808} & \textbf{0.760} \\
\midrule
Min-K & 0.672 & 0.645 \\
Min-K + \ourmethod & \textbf{0.784} & \textbf{0.734} \\
\midrule
Min-K++ & 0.555 & 0.584 \\
Min-K++ + \ourmethod & \textbf{0.565} & \textbf{0.595} \\
\bottomrule
\end{tabular}

\label{tab:deepseek-transfer}
\end{table}

\textbf{Human as Launderer.}
While LLMs enable automated, large-scale rewriting practices, real-world laundering may not be carried out solely by LLMs. We simulate a laundering scenario using the Polite dataset~\citep{wang2022pay}, in which human annotators rewrote impolite sentences into polite versions.
In this experiment, we treat the polite versions as "laundered" training data $D_{\rm train}$, while their original versions are treated as $D_{\rm pro}$. We find that \ourmethod~improves most off-the-shelf detectors (from Table~\ref{tab:polite_sdr}).

\begin{table}[ht]
\caption{Detection performance on the Polite (human-rewritten) dataset. 
\ourmethod~improves most detectors even when laundering is performed by humans, 
demonstrating \ourmethod’s robustness and transferability beyond LLM-based pipelines.}
\centering
\vspace{4pt}\footnotesize
\begin{tabular}{lcc}
\toprule
Method & AUC & ASR \\
\midrule
Recall & \textbf{0.649} & \textbf{0.635} \\
Recall + \textbf{\ourmethod} & 0.566 & 0.575 \\
\midrule
Loss & 0.687 & 0.680 \\
Loss + \textbf{\ourmethod} & \textbf{0.751} & \textbf{0.725} \\
\midrule
Min-K & 0.652 & 0.680 \\
Min-K + \textbf{\ourmethod} & \textbf{0.713} & \textbf{0.690} \\
\midrule
Min-K++ & 0.560 & 0.575 \\
Min-K++ + \textbf{\ourmethod} & \textbf{0.624} & \textbf{0.615} \\
\bottomrule
\end{tabular}

\label{tab:polite_sdr}
\end{table}

We notice that Recall+\ourmethod~becomes slightly worse than Recall alone, which we attribute to polite expressions such as “I think” being already common in pretraining, making both training and non-training data equally easy to continue and collapsing the gap Recall relies on. Other detectors (+\ourmethod) remain generally effective. In practice, Loss (+\ourmethod) and Min-K (+\ourmethod) may provide more reliable signals in this setting.
Overall, these results demonstrate that \ourmethod~remains effective when the laundering pipeline uses a different LLM or human rewriting, and when the auxiliary and laundering models are mismatched, confirming the transferability of \ourmethod~in diverse and practical scenarios.

\section{Mixed-Register Laundering Pipelines}
\label{app:mixed}

We have conducted experiments on \textbf{mixed-register transformations} by combining two registers in the laundering prompt (e.g., ``opinion blog post with a persuasive tone'', ``storytelling narrative as a sports report'', ``informational description like a recipe''). Across all three mixed prompts, \ourmethod~consistently improves detection. For each mixed prompt, we present the recovered prompt followed by the corresponding detection results.

\noindent\textbf{Mixed-register prompt 1.}
Rewrite the text as an \textbf{opinion blog post} with a \textbf{persuasive tone}. 

\noindent\emph{\ourmethod-reversed prompt:} Rewrite the text as an opinion blog post, emphasizing personal narratives and reflections that highlight the emotional, cultural, or historical significance of the subject while engaging the reader.

\noindent\textbf{Mixed-register prompt 2.}
Rewrite the text as a \textbf{storytelling narrative} to introduce information as a \textbf{sports report}.  

\noindent\emph{\ourmethod -reversed prompt:} Rewrite the text as a narrative blog post. Focus on transforming the original information into a more engaging story, emphasizing the historical context and the evolution of the organization while maintaining a conversational tone.

\noindent\textbf{Mixed-register prompt 3.}
Rewrite the text as an \textbf{informational description} like a \textbf{recipe}.  

\noindent\emph{\ourmethod-reversed prompt:} Rewrite the text as a step-by-step instructional guide, clearly outlining key aspects and maintaining a logical flow and numbered steps.

\begin{table}[H]
\centering
\caption{\ourmethod~transferability on mixed-register laundering prompts. \ourmethod~consistently improves AUC and ASR across three mixed-register laundering transformations, demonstrating robustness to more complex combinations of stylistic rewrites.}
\vspace{4pt}\footnotesize
\setlength{\tabcolsep}{5.5pt}
\begin{tabular}{lcc|cc|cc}
\toprule
\multirow{2}{*}{Method}  &
\multicolumn{2}{c}{Mixed Prompt 1} &
\multicolumn{2}{c}{Mixed Prompt 2} &
\multicolumn{2}{c}{Mixed Prompt 3} \\
\cmidrule(lr){2-3} \cmidrule(lr){4-5} \cmidrule(lr){6-7}
 & AUC & ASR & AUC & ASR & AUC & ASR \\
\midrule
Recall            & 0.559  & 0.560  & 0.666 & 0.650 & 0.737 & 0.715 \\
Recall + \ourmethod      & \textbf{0.757} & \textbf{0.705} &
                     \textbf{0.777} & \textbf{0.725} &
                     \textbf{0.831} & \textbf{0.770} \\[2pt]
                     \midrule
Loss              & 0.572  & 0.575  & 0.670 & 0.655 & 0.738 & 0.685 \\
Loss + \ourmethod         & \textbf{0.767} & \textbf{0.705} &
                     \textbf{0.796} & \textbf{0.750} &
                     \textbf{0.834} & \textbf{0.785} \\[2pt]
                     \midrule
Min-K             & 0.587  & 0.580  & 0.677 & 0.670 & 0.747 & 0.705 \\
Min-K + \ourmethod        & \textbf{0.750} & \textbf{0.715} &
                     \textbf{0.749} & \textbf{0.700} &
                     \textbf{0.820} & \textbf{0.765} \\[2pt]
                     \midrule
Min-K++           & 0.475  & 0.535  & 0.475 & 0.515 & 0.586 & 0.605 \\
Min-K++ + \ourmethod      & \textbf{0.586} & \textbf{0.615} &
                     \textbf{0.538} & \textbf{0.580} &
                     \textbf{0.617} & \textbf{0.615} \\
\bottomrule
\end{tabular}

\label{tab:mixed_register}
\end{table}

Across all three mixed-register prompts, \ourmethod~consistently improves both AUC and ASR, demonstrating that it can successfully transfer to mixed laundering transformations. Nevertheless, we agree that extremely exotic transformations (e.g., pseudo-translation into low-resource languages) may further stress \ourmethod.

\section{Partial Laundering Experiment}
\label{app:partial}
We conducted an experiment to evaluate whether \ourmethod~remains effective when only a portion of the proprietary data is laundered. 
Given $\lvert D_{\text{pro}} \rvert = 200$, we randomly selected half of the samples and applied the first five inside-register prompts (Table~\ref{tab:inside}) to generate the laundered subsets used for training, while the remaining samples were left unaltered. 
\ourmethod~was then applied to the full $D_{\text{pro}}$ to infer the laundering transformation. 
Table~\ref{tab:par} below reports the average performance across all five prompts.
\ourmethod~successfully improves detection performance across all four off-the-shelf detectors. For example, the AUC/ASR of Loss increases from 0.599/0.597 to 0.732/0.703.

\begin{table}[H]
\centering
\caption{Average performance across all five laundering prompts under partial laundering.}
\label{tab:par}
\vspace{4pt}\footnotesize
\begin{tabular}{lcc}
\toprule
Method & Avg AUC & Avg ASR \\
\midrule
Recall            & 0.608 & 0.590 \\
Recall + \ourmethod~     & \textbf{0.722} & \textbf{0.694} \\
\midrule
Loss              & 0.599 & 0.597 \\
Loss + \ourmethod~       & \textbf{0.732} & \textbf{0.703} \\
\midrule
Min-K             & 0.608 & 0.603 \\
Min-K + \ourmethod~      & \textbf{0.724} & \textbf{0.701} \\
\midrule
Min-K++           & 0.553 & 0.565 \\
Min-K++ + \ourmethod~    & \textbf{0.580} & \textbf{0.602} \\
\bottomrule
\end{tabular}
\end{table}

\section{Hyperparameter Sensitivity Study}
\label{app:hyper}
We conducted an extended sensitivity study to examine how \ourmethod~behaves under different choices of the key hyper-parameters $K$ (see Table~\ref{K}), $l$ (see Table~\ref{l}), $m$ (see Table~\ref{m}), and $n$ (see Table~\ref{n}), using Loss+\ourmethod~under the first inside-register prompt in Table~\ref{tab:inside}.

Our key findings are as follows:
Increasing $K$ from $3 \to 15$ improves AUC from $0.71 \to 0.75$ and ASR from $0.69 \to 0.71$, at the cost of higher runtime and query budget (roughly \$7--\$39 with GPT-4o; substantially cheaper with GPT-4o-mini). The value of $m$ has the strongest impact. Varying $m$ from $3 \to 9$ raises AUC from $0.72 \to 0.81$ and ASR from $0.69 \to 0.76$ with almost unchanged query cost, making $m=9$ a strong default choice. $l$ and $n$ exhibit moderate gains. Setting $l=7$ and $n=7$ achieves a good balance between performance and cost. \emph{Note that $n$ is used only once when building templates and does not affect per-run latency.}

\noindent\textbf{Detailed numerical results.}  
Unless otherwise specified, we use $K=10$, $l=5$, $m=5$, and $n=10$ as the default hyper-parameters.
\begin{table}[H]
\centering
\caption{\protect Sensitivity to $K$.}
\label{K}
\vspace{4pt}\footnotesize
\begin{tabular}{c|c|c|c|c}
\toprule
$K$ & AUC & ASR & Time (hh:mm:ss) & Query Budget (GPT4o / mini) \\
\midrule
3  & 0.712 & 0.685 & 02:23:42 & \(\sim\$7 / 0.8\) \\
5  & 0.727 & 0.690 & 04:11:33 & \(\sim\$13 / 1.5\) \\
15 & 0.747 & 0.705 & 13:27:12 & \(\sim\$39 / 4.5\) \\
\bottomrule
\end{tabular}
\end{table}

\begin{table}[H]
\centering
\caption{Sensitivity to $l$.}
\label{l}
\vspace{4pt}\footnotesize
\begin{tabular}{c|c|c|c|c}
\toprule
$l$ & AUC & ASR & Time (hh:mm:ss) & Query Budget (GPT4o / mini) \\
\midrule
3 & 0.715 & 0.670 & 08:21:33 & \(\sim\$10 / 1\) \\
7 & 0.731 & 0.695 & 09:13:12 & \(\sim\$10 / 1\) \\
9 & 0.718 & 0.695 & 09:35:17 & \(\sim\$10 / 1\) \\
\bottomrule
\end{tabular}
\end{table}

\begin{table}[H]
\centering
\caption{Sensitivity to $m$.}
\label{m}
\vspace{4pt}\footnotesize
\begin{tabular}{c|c|c|c|c}
\toprule
$m$ & AUC & ASR & Time (hh:mm:ss) & Query Budget (GPT4o / mini) \\
\midrule
3  & 0.718 & 0.685 & 08:14:33 & \(\sim\$10 / 1\) \\
7  & 0.796 & 0.740 & 08:19:29 & \(\sim\$10 / 1\) \\
9  & 0.810 & 0.760 & 08:26:45 & \(\sim\$10 / 1\) \\
11 & 0.806 & 0.755 & 08:33:34 & \(\sim\$10 / 1\) \\
\bottomrule
\end{tabular}
\end{table}

\begin{table}[H]
\centering
\caption{Sensitivity to $n$. $n$ controls the construction of the template and is computed only once during initialization. Therefore, it does not contribute to the runtime or query budget.}
\label{n}
\vspace{4pt}\footnotesize
\begin{tabular}{c|c|c|c|c}
\toprule
$n$ & AUC & ASR & Time (hh:mm:ss) & Query Budget \\
\midrule
3 & 0.705 & 0.675 & -- & -- \\
7 & 0.752 & 0.705 & -- & -- \\
9 & 0.742 & 0.695 & -- & -- \\
\bottomrule
\end{tabular}
\end{table}
Overall, a practical and efficient configuration for \ourmethod~is:
$(K=5,\quad l=7,\quad m=9,\quad n=7)$,
which attains most of the performance gains at moderate cost. 
% We will include these tables and a concise discussion in the Appendix.

\section{Comparison to Reverse Prompt Engineering Methods}
\label{app:reverse}
To clarify how \ourmethod~differs from related lines of work, especially those based on prompt-recovery techniques, we note an important methodological distinction:
To the best of our knowledge, existing \emph{reverse prompt engineering} methods, however, assume that the unknown prompt is applied at \emph{inference} time and leaves a direct trace in the observed outputs. In our setting, the laundering prompt is applied before \emph{training}. The data-rights holder only sees the trained model and never observes any text generated under the unknown laundering prompt, making a direct application impossible.

We summarize the conceptual differences between Reverse Prompt Engineering~\citep{li2024reverse} and \ourmethod~as follows. Reverse prompt engineering assumes that the unknown prompt is applied during inference, meaning that it directly conditions the model’s outputs and therefore leaves observable traces that the recovery procedure can exploit. In contrast, \ourmethod~targets laundering transformations applied before training, where the prompt does not manifest in any generated text. Instead, only implicit consequences of the laundering process remain embedded in the trained model. Moreover, reverse prompt engineering methods rely on having direct access to model outputs produced under the unknown prompt, whereas in our setting, the data-rights holder never observes any prompt-conditioned outputs and can only query the final trained model. These differences make existing reverse prompt engineering approaches incompatible with the laundering-before-training scenario that \ourmethod~is designed to address.

To still provide a comparison, we slightly adapt Reverse Prompt Engineering~\citep{li2024reverse} to a use case better aligned with our problem setup: given an original document, we let $M_t$ generate a continuation and treat that continuation as a proxy “laundered’’ version; we then ask~Reverse Prompt Engineering~\citep{li2024reverse} to infer a prompt connecting the original and continuation.

Table~\ref{tab:reverse} shows that \ourmethod~provides the largest improvement among all methods. Loss+\ourmethod\ achieves an AUC of 0.782 and an ASR of 0.764, clearly outperforming both the baseline detector and the adapted reverse prompt engineering approach. While reverse prompt engineering offers modest gains over the baseline, its improvements remain limited, highlighting that it is considerably less effective than \ourmethod~in recovering laundering transformations and restoring detector performance.

\begin{table}[!t]
\caption{Comparison between \ourmethod~and reverse prompt engineering.}
\label{tab:reverse}
\centering
\vspace{4pt}\footnotesize
\begin{tabular}{lcc}
\toprule
Method & AUC & ASR \\
\midrule
Loss & 0.638 & 0.635 \\
Loss + Reverse Prompt Engineering & 0.682 & 0.663 \\
\textbf{Loss + \ourmethod} & \textbf{0.782} & \textbf{0.764} \\
\bottomrule
\end{tabular}
\end{table}

\section{TPR@1\% Result}
\label{app:tpr}
We report TPR@1\% (True Positive Rate at 1\% False Positive Rate) 
as a complementary metric to TPR@5\% reported in the main paper. While TPR@5\% evaluates detection performance under a relatively lenient false-positive constraint, TPR@1\% imposes a stricter threshold, reflecting scenarios where auditors require higher precision before accusing of data misuse. A higher TPR@1\% indicates that the detection method can correctly identify more training samples 
while maintaining a very low rate of false accusations. As shown in 
Table~\ref{tab:tpr1}, \ourmethod~consistently improves TPR@1\% across all 
detectors and both inside- and outside-register settings, further confirming its 
effectiveness under stricter evaluation criteria.

\begin{table}[!t]
\centering
\caption{TPR@1\% results for the main experiment.}
\label{tab:tpr1}
\vspace{4pt}\footnotesize
\begin{tabular}{lc|c}
\toprule
Method & Inside TPR@1\% & Outside TPR@1\% \\
\midrule
Recall                & 0.002  & 0.008 \\
Recall + \ourmethod   & \textbf{0.106} & \textbf{0.062} \\
\midrule
Loss                  & 0.047  & 0.000 \\
Loss + \ourmethod     & \textbf{0.112} & \textbf{0.069} \\
\midrule
Min-K                 & 0.001  & 0.016 \\
Min-K + \ourmethod    & \textbf{0.096} & \textbf{0.082} \\
\midrule
Min-K++               & 0.002  & 0.012 \\
Min-K++ + \ourmethod  & \textbf{0.086} & \textbf{0.071} \\
\bottomrule
\end{tabular}
\end{table}

\section{Limitation}
\label{limitaion}
A key limitation of our current approach is that the register taxonomy it relies on is too coarse-grained to locate goals accurately. Although the existing taxonomy of 23 sub-registers offers broad coverage of textual styles, it was not initially designed for classifying the laundering goal. Consequently, there are cases where none of the 23 registers can adequately capture the intent of a synthesized prompt, leading to reduced accuracy in goal identification and, in turn, lower quality in restored prompts. Overcoming this limitation calls for future research on developing more fine-grained taxonomies tailored to synthesized data, thereby enabling more accurate and robust prompt reversal in practical scenarios.

\section{Ethics Statement}
This work focuses on defending against unauthorized data laundering in LLM training by restoring the effectiveness of post-hoc detection. All experiments were conducted on publicly available datasets, and no proprietary or personal data was used. While there is a risk that our techniques could be misused to improve laundering attacks, our contributions are explicitly framed for defensive purposes, aiming to strengthen accountability and responsible governance in AI systems.

\section{Reproducibility Statement}
We have made significant efforts to ensure the reproducibility of our results. All datasets used in this paper are publicly available, and details of data synthesis procedures are provided in Appendix~\ref{app:prompts}. Complete descriptions of model architectures, training settings, and evaluation protocols are included in the main paper and Appendix~\ref{sec:detail implement}. For each experiment, we specify hyper-parameters, implementation details, and the auxiliary LLM prompts used for synthesis and reverse synthesis. Our codebase, built on PyTorch and Hugging Face Transformers, was included in the submitted supplementary material and will be released upon publication to facilitate full replication of our experiments.

\section{AI usage Clarification}
Large Language Models improved the manuscript’s grammar and readability; all research design, analysis, and interpretation were conducted by the authors.

% \ourmethod~successfully improves detection performance across all four off-the-shelf detectors. 
% For example, the AUC/ASR of Loss increases from 0.599/0.597 to 0.732/0.703.